\documentclass[
 reprint,
 amsmath,amssymb,
 aps,
 10pt,
 nofootinbib
 longbibliography,
 superscriptaddress
 ]{revtex4-2}

\usepackage{graphicx}
\usepackage{dcolumn}
\usepackage{bm}
\usepackage[utf8]{inputenc}

\DeclareMathOperator{\RX}{RX}
\newcommand{\ket}[1]{| #1\rangle}
\newcommand{\dc}{\mathrm{dc}}
\newcommand{\ac}{\mathrm{ac}}

\usepackage[urlcolor=blue, citecolor=magenta, linkcolor=magenta, hyperindex, colorlinks, bookmarks=true]{hyperref} 

\begin{document}

\title{Dynamical sweet spot engineering via two-tone flux modulation of superconducting qubits}
\author{Joseph A. Valery}\email[Corresponding author: ]{joseph@rigetti.com}
\affiliation{Rigetti Computing, 775 Heinz Avenue, Berkeley, CA, USA 94710}
\author{Shoumik Chowdhury}
\affiliation{Rigetti Computing, 775 Heinz Avenue, Berkeley, CA, USA 94710}
\affiliation{Department of Physics, Yale University, New Haven, CT, USA 06520}
\author{Glenn Jones}
\affiliation{Rigetti Computing, 775 Heinz Avenue, Berkeley, CA, USA 94710}
\author{Nicolas Didier}
\affiliation{Rigetti Computing, 775 Heinz Avenue, Berkeley, CA, USA 94710}

\date{\today}

\begin{abstract}

Current superconducting quantum processors require strategies for coping with material defects and imperfect parameter targeting in order to scale up while maintaining high performance. To that end, in-situ control of qubit frequencies with magnetic flux can be used to avoid spurious resonances. However, increased dephasing due to $1 / f$ flux noise limits performance at all of these operating points except for noise-protected sweet spots, which are sparse under DC flux bias and monochromatic flux modulation. Here we experimentally demonstrate that two-tone flux modulation can be used to create a continuum of dynamical sweet spots, greatly expanding the range of qubit frequencies achievable while first-order insensitive to slow flux noise. To illustrate some advantages of this flexibility, we use bichromatic flux control to reduce the error rates and gate times of parametric entangling operations between transmons. Independent of gate scheme, the ability to use flux control to freely select qubit frequencies while maintaining qubit coherence represents an important step forward in the robustness and scalability of near-term superconducting qubit devices.

\end{abstract}

\maketitle

\section{Introduction}

As they scale, superconducting quantum computing architectures need the capability to contend with experimental imperfections. Two-level system defects~\cite{Martinis_2005,Muller_2015,Klimov:2018,Burnett:2019,Schlor:2019,Lisenfeld:2019,Bilmes:2019} and imperfect frequency targeting due to variance in Josephson junction fabrication~\cite{Lehnert_1992,Oliva_1994,Potts_2001,Koppinen_2007,Granata_2008,Bumble_2009,Pop_2012,Costache_2012,Wu_2017,Brink:2018,Kreikebaum_2020} are common sources of error on today's processors. Taking advantage of modular device compositions~\cite{Dickel:2018,Chou:2018,Zhong:2020,Gold_2021} can help avoid these problems, but it is also necessary to use in-situ control to cope with them as they arise. For example, the use of parallel Josephson junctions in a SQUID loop allows for the tunability of qubit frequency with magnetic flux, providing a convenient way to correct for variations from frequency targets as well as dodge coupled material defects. Additional pathways for control necessarily open up additional pathways for noise, however, and the cost of flux tunability is exposure to $1/f$ flux noise, which dephases tunable qubits~\cite{Koch:1983, Wellstood:1987,Vion:2002,Martinis:2003,Ithier:2005,Yoshihara:2006, Bialczak:2007,KochR:2007,Faoro:2008,Manucharyan:2009,Barends:2013,Omalley:2015,Kumar:2016,Yan:2016,Plourde:2017,Kou:2017,Quintana:2017,You:2019}. Thus, it is important to develop control schemes that minimize the time a qubit spends at flux-sensitive operating points during a program~\cite{Vion:2002,Rol:2019,Huang:2020}. 

Under a DC flux bias $\Phi_{\dc}$ and a single-tone flux modulation with amplitude $\Phi_{\ac}$, flux-insensitive ``sweet spot'' operating points are sparse. There are six within a single flux period $\Phi_0$, corresponding to $\Phi_{\dc} = 0, 0.5\Phi_0$ along with $\Phi_{\ac} \approx 0, 0.6\Phi_0$, as well as $\Phi_{\dc} = \pm 0.25\Phi_0$ along with $\Phi_{\ac} \approx 0.4\Phi_0$ ~\cite{Nico:2019}. The result is only six time-averaged qubit frequencies $\bar{f}$ that are achievable without exposing the qubit to flux noise-induced dephasing.  However, a recent theory proposal suggests that the introduction of a second flux modulation tone can provide a pathway to achieving more flexibility~\cite{Nico:bichro}. Whereas a DC flux bias provides two sweet spots per flux period and a single tone of modulation introduces four more, bichromatic modulation unlocks a continuum of operating points that are first-order insensitive to $1/f$ flux noise. Control of the frequency relationship, mixing, and relative phase of the two tones can significantly shift a tunable qubit's average frequency under modulation, creating dynamical sweet spots across a range of $\bar{f}$. 

We experimentally demonstrate this control in Sec.~\ref{sweet_spots}. After briefly introducing the bichromatic pulses and superconducting circuit devices used in this study, we show that protection from flux noise-induced dephasing can be achieved across a significant portion of the qubit tunability band. In Sec.~\ref{gates}, we use the flexibility offered by bichromatic flux modulation to improve the performance of parametric entangling gates~\cite{Bertet:2006,Niskanen:2007,Rigetti:2009,Beaudoin:2012,Strand:2013,McKay:2016,Naik:2017,Roth:2017,Mundada:2019}. During a parametric gate, the frequency of a tunable qubit is modulated by an AC flux pulse to produce sidebands at multiples of the modulation frequency, centered around the time-averaged frequency $\bar{f}$. The use of these sidebands to activate entangling interactions with a neighboring qubit means that $\bar{f}$ can be freely chosen, as the frequency of modulation can be adjusted to maintain  resonance~\cite{Matt:2018, Shane:2018}. As a result, gates can be operated at dynamical sweet spots and the average frequencies to which they correspond~\cite{Sab:2019, Deanna_2020, Schuyler:2019, Nico:2019}. We demonstrate that bichromatic modulation can greatly increase the robustness and scalability of this gate scheme by enabling precise control of $\bar{f}$ at those dynamical sweet spots, dodging frequency collisions. Furthermore, in Sec.~\ref{sidebands} we demonstrate that bichromatic modulation can be used to reduce the duration of parametric gates. The freedom  provided by the use of sidebands to perform these gates comes at the cost of lower effective qubit-qubit coupling, which is renormalized across all generated sidebands~\cite{Nico:2018}. By presenting a way to alter the time-dependence of a tunable qubit's frequency during modulation, bichromatic control can be used to distribute coupling more-optimally across sidebands and speed up specific interactions.

\section{Dynamical Sweet Spots}
\label{sweet_spots}

Here we set out to demonstrate the effectiveness of bichromatic flux modulation in achieving protection against $1 / f$ flux noise across a wide bandwidth of time-averaged qubit frequencies $\bar{f}$. 
We employ the following convention in parameterizing an applied flux $\Phi(t)$ involving a DC bias and two modulation frequencies:
\begin{equation}
    \begin{split}
        \Phi(t) = \Phi_{\dc} + \Phi_{\ac}u(t)\Big{[}&\cos(\alpha)\cos\big{(}2{\pi}f_mt\big{)}\\+ \text{ }&\sin(\alpha)\cos\big{(}2{\pi}pf_mt + \theta\big{)}\Big{]}.
    \end{split}
    \label{general_waveform}
\end{equation}
In this expression, $\Phi_{\dc}$ is a DC bias amplitude, $\Phi_{\ac}$ is a modulation amplitude, $u(t)$ is a waveform envelope scaled to unity, $\alpha$ is a mixing angle defining the relative strength of each tone, $f_m$ is the fundamental frequency of modulation, $p$ is a multiplier on $f_m$ that determines the second modulation tone, and $\theta$ is a relative phase between the tones~\cite{Nico:bichro}. The envelope $u(t)$ is implemented as a flat-top pulse with a symmetric rise and fall described by the Gauss error function.   

\begin{figure}
    \includegraphics[width=\columnwidth]{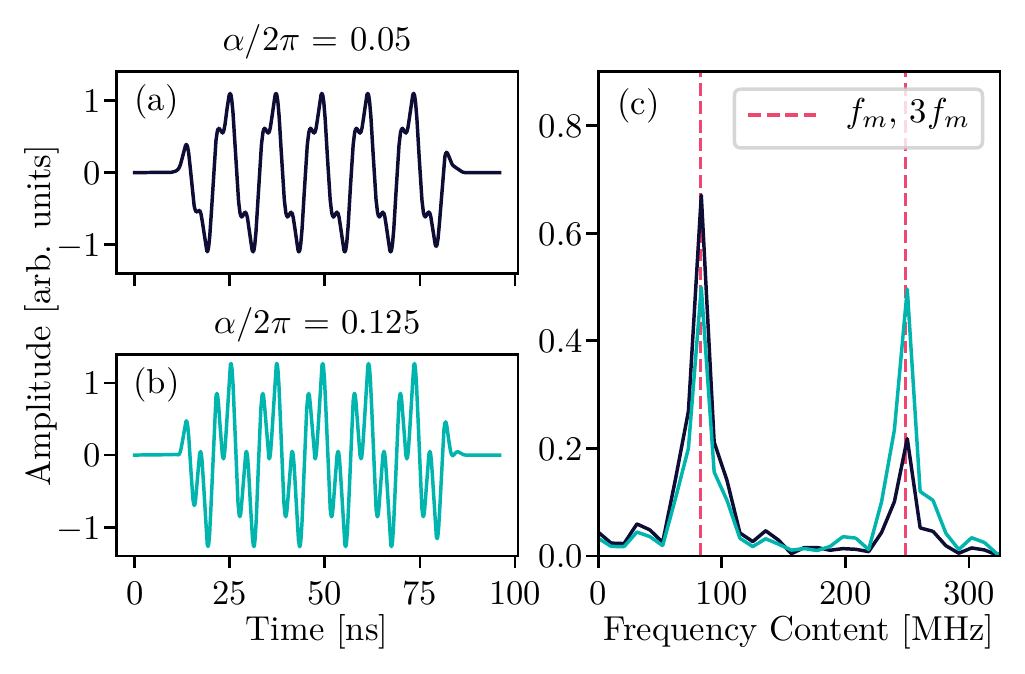}
    \caption{Left panels: Example waveforms described by Eq.~\eqref{general_waveform}. The pulses in (a) and (b) differ only in mixing angle $\alpha$, otherwise sharing the following parameter values: $p=3$, $f_m = 82.9\,\mathrm{MHz}$, $\theta/2\pi = 0.675$. Right panel (c): The Fourier transform of the pulses in (a) and (b), showing content at $f_m$ and $3f_m$ with relative strengths dependent on $\alpha$. For $\alpha/2\pi = 0.125$, displayed in teal, both tones are weighted equally. Monochromatic modulation is produced at $\alpha/2\pi = 0 \text{ and } 0.25$ (not shown) with frequency $f_m$ and $3f_m$ respectively.}    
    \label{fig:pulse_shape}
\end{figure}
Two examples of bichromatic pulses defined in this way are shown in Fig.~\ref{fig:pulse_shape} along with their Fourier transforms, showing peaks in frequency content at $f_m$ and $pf_m$ with amplitudes according to $\alpha$. In-depth information regarding the generation of this class of pulses and the calibration of specific $\alpha$ and $\theta$ values is provided in Appendix~\ref{pulse_calibration}.

\begin{figure}
    \includegraphics[width=\columnwidth]{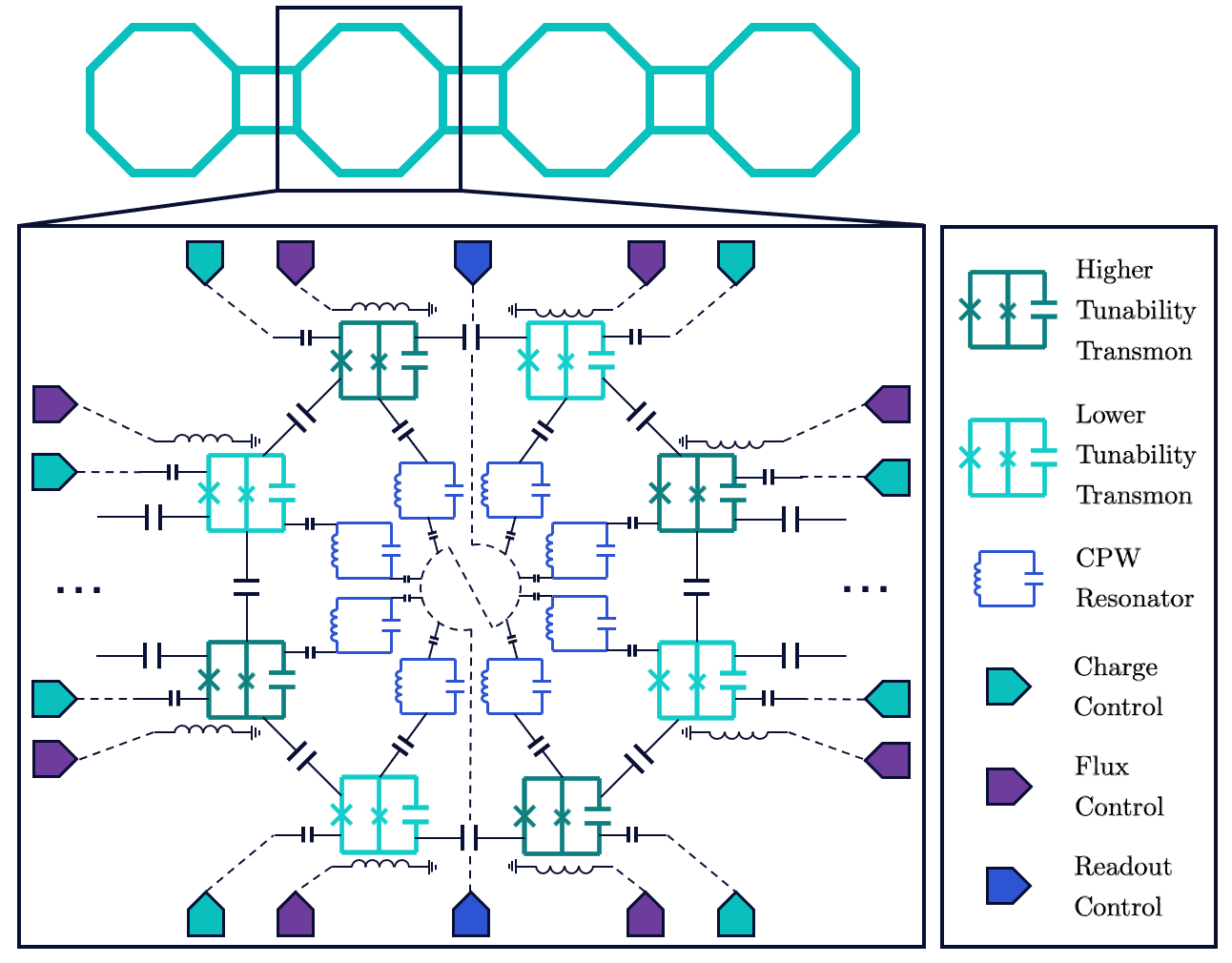}
    \caption{A circuit diagram representation of the devices used in this study. The total lattice includes 32 tunable transmons, with the topology of eight presented here constituting an octagonal unit cell that is repeated four times to make up a full device. Dispersive readout is achieved via coplanar waveguide resonators capacitively coupled to each transmon, and readout signal delivery is multiplexed eight-fold by octagon. These transmons are tunable via inductively coupled flux lines and excitable via capacitively coupled RF drive lines. There is a static capacitive coupling between each pair of neighboring qubits. The lattice alternates between two classes of transmon that differ in frequency band and tunability. Parametric gates are performed by modulating the higher frequency, higher tunability qubits and placing negative-indexed sidebands on resonance with the capacitively-coupled, lower-frequency neighbors.}
    \label{fig:circuit_diagram}
\end{figure}

The experiments described in this paper are performed on superconducting transmons~\cite{Koch:2007} within two 32-qubit Rigetti quantum processors. Both of these devices are described by the circuit diagram in Fig.~\ref{fig:circuit_diagram} and share the properties listed in its caption. Relevant information about the specific qubits under study can be found in Table~\ref{qubit_table}.    

\begin{table*}
\begin{ruledtabular}
\begin{tabular}{r | c c c c c c c}
& $f_{01}$ [GHz] & Tunability [$\mathrm{GHz}$] & Anharmonicity [GHz] & $T_1$ [$\mu$s] & $T_2^*$ [$\mu$s] & Readout Fidelity [\%] & 1Q Gate Fidelity [\%] \\
\hline
Qubit 1 & 5.250 & 0.824 & -0.205 & 33 $\pm$ 6 & 46 $\pm$ 4 & 96.07 $\pm$ 2.80 & 99.89 $\pm$ 0.01 \\
Qubit 2 & 4.269 & 0.401 & -0.187 & 25 $\pm$ 2 & 23 $\pm$ 4 & 94.24 $\pm$ 0.72 & 99.92 $\pm$ 0.02 \\
\hline
Qubit 3 & 4.791 & 1.074 & -0.206 & 25 $\pm$ 6 & 22 $\pm$ 2 & 97.02 $\pm$ 0.81 & 99.83 $\pm$ 0.02  \\
Qubit 4 & 3.365 & 0.170 & -0.201 & 36 $\pm$ 8 & 20 $\pm$ 4 & 92.72 $\pm$ 2.57 & 99.85 $\pm$ 0.01  \\

\end{tabular}
\end{ruledtabular}
\caption{Relevant information about the qubits used in the experiments described in this manuscript. Qubits 1 and 2 are capacitively-coupled neighbors, as are qubits 3 and 4 on a different device. The indexing labels use here are for convenience and are not representative of positioning within a lattice. The listed $f_{01}$ values correspond to bias $\Phi_{\dc} = 0$, and represent the maximum of each qubit's tunability band. Listed anharmonicities correspond to $f_{12} - f_{01}$ at $\Phi_{\dc} = 0$. Listed $T_1$ and $T_2^*$ values represent coherence times measured with $\Phi_{\dc} = \Phi_{\ac} = 0$, when qubits are biased at their maximum frequencies and are not being modulated. Listed readout fidelities represent the average of classification probabilities $P(0|0)$ and $P(1|1)$ measured non-simultaneously across qubits. Listed 1Q gate fidelities are measured with randomized benchmarking, performed non-simultaneously across qubits.}
\label{qubit_table}
\end{table*}

To first order, the dephasing rate $\Gamma_{\phi}$ is proportional to the following:
\begin{equation}
    \Gamma_\phi \propto \sqrt{A_{\dc}^2\left(\frac{\partial\bar{f}}{\partial\Phi_{\dc}}\right)^2 + A_{\ac}^2\left(\frac{\partial\bar{f}}{\partial\Phi_{\ac}}\right)^2},
\end{equation}
where $A_{\dc}$ and $A_{\ac}$ are the strengths of additive and multiplicative $1/f$ noise respectively. For experimental simplicity we track the transverse relaxation rate $\Gamma_2 = \Gamma_1/2 + \Gamma_{\phi}$, and transverse relaxation time $T_2^* = 1/\Gamma_2$, in lieu of $\Gamma_{\phi}$ and $T_{\phi} = 1/\Gamma_{\phi}$, as $\Gamma_1$ exhibits a comparably weak dependence on slow flux noise~\cite{Krantz_2019,Sab:2019}. 

Choosing $p$, the frequency multiplier of the second bichromatic tone, to be an odd, positive integer ensures that $\partial \bar{f}/\partial \Phi_{\dc}$ is 0 at the same biases where it is 0 without modulation~\cite{Nico:bichro}. That is, for a tunable qubit with flux period $\Phi_0$, DC sweet spots persist at $\Phi_{\dc} = 0 \text{ and } \frac{1}{2}\Phi_0$ regardless of $\Phi_{\ac}$. As a result, the DC flux bias can remain constant throughout a pulse program without contributing to increased dephasing during bichromatic modulation. Another advantage of maintaining these DC sweet spots in particular is that they occur at biases around which the qubit frequency is symmetric. This symmetry eliminates odd-indexed sidebands that would otherwise be generated during modulation, cleaning the spectrum for parametric gates and leaving more renormalized coupling to be distributed amongst the remaining even sidebands (see Appendix~\ref{theory}). To realize these benefits, all qubits in this study are biased at $\Phi_{\dc} = 0$ and all bichromatic modulation uses a frequency multiplier of $p=3$.

With $\partial \bar{f}/\partial \Phi_{\dc}$ equal to 0 throughout operation, we expect $T_2^*$ to track specifically with $\partial \bar{f}/\partial \Phi_{\ac}$. At a consistent $\Phi_{\dc}$ of $0$, the only non-zero amplitude of monochromatic modulation in the first flux period for which $\partial \bar{f}/\partial \Phi_{\ac} = 0$ is $\Phi_{\ac} \approx 0.6\Phi_0$, corresponding to a minimum in $\bar{f}$.
\begin{figure}
    \includegraphics[width=\columnwidth]{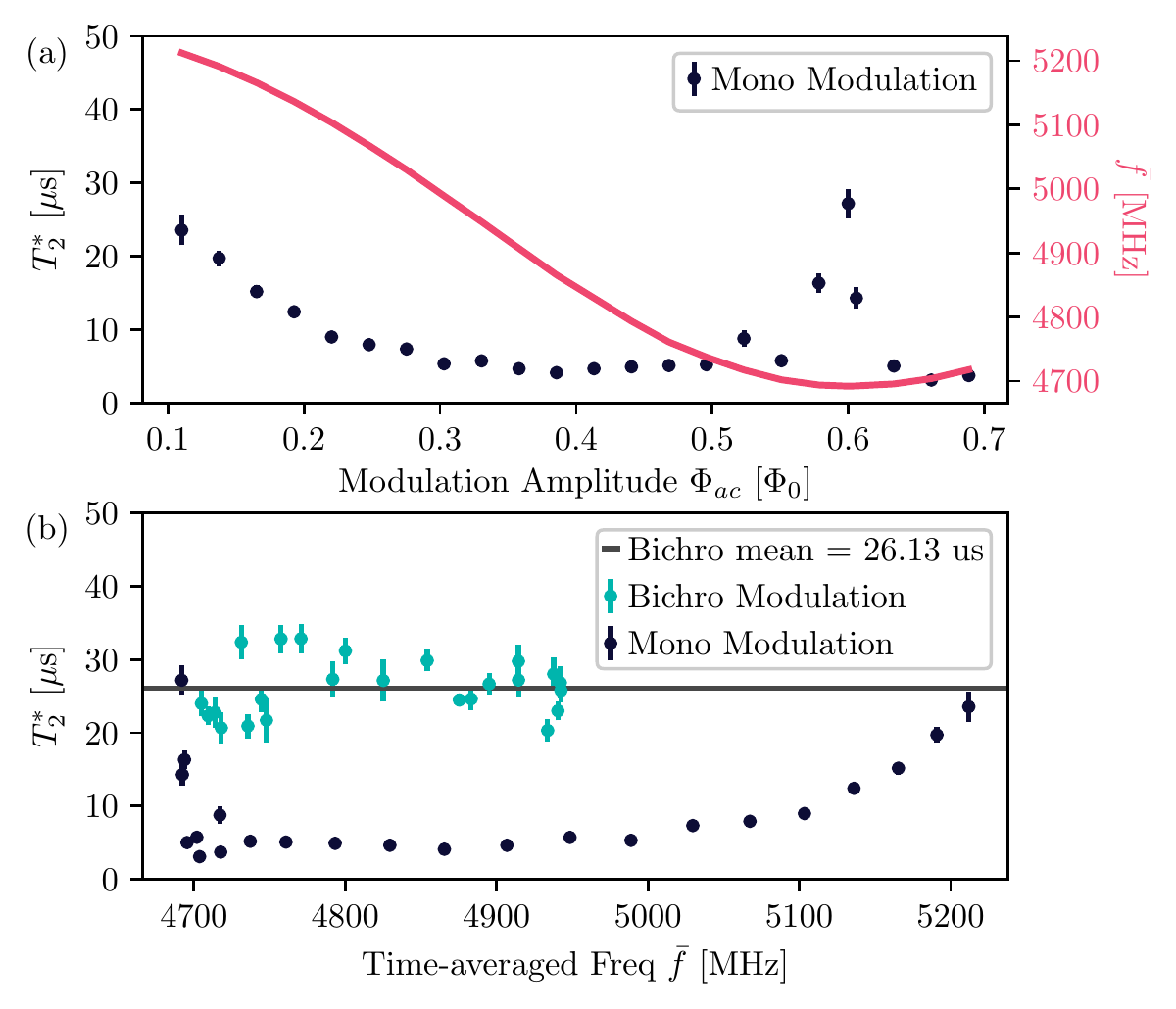}
    \caption{(a) Measured, time-averaged frequency $\bar{f}$ and $T_2^*$ of qubit 1 during monochromatic modulation at $f_m=100\,\mathrm{MHz}$, plotted as a function of modulation amplitude. $T_2^*$ falls significantly at amplitudes for which $\bar{f}$ is flux-sensitive, only recovering at $0.6\Phi_0$  where $\partial \bar{f}/\partial \Phi_{\ac} = 0$. Each data point is the average of four $T_2^*$ measurements, weighted by their relative error. (b) The same $T_2^*$ during monochromatic modulation data, plotted against $\bar{f}$ instead of modulation amplitude. The $T_2^*$ times in teal are measured specifically at the sweet spots produced by bichromatic modulation with a fundamental frequency of $f_m=100\,\mathrm{MHz}$ and different values of mixing angle $\alpha$ and relative phase $\theta$. The $T_2^*$ resurgence achieved at the single monochromatic sweet spot is maintainable, with at least the demonstrated operating point density, across a spectrum of bichromatic sweet spots corresponding to approximately $240\,\mathrm{MHz}$ of $\bar{f}$ flexibility.}    
    \label{fig:coherence}
\end{figure}
This is shown in Fig.~\ref{fig:coherence}(a), where we measure $\bar{f}$ and $T_2^*$ of qubit 1 (ref.~Table~\ref{qubit_table}) as a function of  $\Phi_{\ac}$ for a fixed, monochromatic modulation frequency of $100\,\mathrm{MHz}$. $\bar{f}$ is measured with a standard Ramsey experiment by playing each flux pulse during the intervening delays between $\RX(\pi/2)$ rotations. Most modulation amplitudes cause $T_2^*$ to drop significantly, with resurgent values only measured around $\Phi_{\ac} = 0.6\Phi_0$. Even there, $T_2^*$ is only around 60\% of the unmodulated value listed in Table~\ref{qubit_table} (strength of $T_2^*$ resurgence is the subject of ongoing research~\cite{Sab:2019,Schuyler:2019}).  

The additional control introduced by bichromatic modulation loosens this operating point restriction significantly. By choosing different mixing angles $\alpha$ and relative phases $\theta$, the $\partial \bar{f}/\partial \Phi_{\ac} = 0$ condition can be met along a continuum of $\bar{f}$ values within the modulated transmon's tunability band~\cite{Nico:bichro}. The corresponding sweet spot $\Phi_{\ac}$ values are no longer fixed to $0.6\Phi_0$. We experimentally demonstrate this flexibility in Fig.~\ref{fig:coherence}(b) by measuring $T_2^*$ specifically at the dynamical sweet spots produced by bichromatic modulation characterized by different pairs of $\alpha$ and $\theta$. It is possible to extend the monochromatic level of $T_2^*$ resurgence across a significant portion of the tunability band, achieving an average $T_2^*$ of $26.13\,\mathrm{{\mu}s}$ where there is a ceiling of around $5\,\mathrm{{\mu}s}$ under monochromatic modulation. 

Not every set of bichromatic control parameters is guaranteed to produce a sweet spot with good $T_2^*$ resurgence, as other sources of decoherence can affect the modulated transmon. The displayed coherence times correspond to a subset of measured operating points at which resurgence was achieved. This down-selection was simply performed with the intention of showing that $T_2^*$-resurgent sweet spots are achievable across this range of $\bar{f}$ with at least the demonstrated density. We attribute the variation in the presented $T_2^*$ values to the defect landscape and temporal fluctuations.

\section{Flexible Gate Operating Points}
\label{gates}

By enabling higher coherence times across a continuous band of time-averaged qubit frequencies during modulation, bichromatic flux pulses can be used to perform higher-fidelity entangling gates across that same band. The freedom to select a specific $\bar{f}$ for a modulated qubit is advantageous for dodging collisions in two ways. First and most directly, any unintended coupling with two-level system (TLS) defects~\cite{Martinis_2005,Muller_2015,Klimov:2018,Burnett:2019,Schlor:2019,Lisenfeld:2019,Bilmes:2019} or other qubits near $\bar{f}$ at the monochromatic sweet spot can be avoided. Second, $\bar{f}$ can be chosen to avoid similar spurious interactions involving any of the generated sidebands. During modulation, these sidebands are produced at frequencies
\begin{equation}
    f_k = \bar{f} + kf_m,
    \label{sideband_freqs}
\end{equation}
where $k$ is the sideband index and $f_m$ is the frequency of modulation. The fixed, monochromatic sweet spot $\bar{f}$ also fixes the modulation frequency required to place a given sideband on resonance with a neighboring qubit to activate an entangling interaction. In the event that any other sidebands generated with this modulation frequency collide with TLS or the resonance condition of a different gate, population is lost from the intended exchange and fidelity drops accordingly. While the modulation frequency is still fixed for a given $\bar{f}$ in the bichromatic case, the value of $\bar{f}$ can be tuned to ensure collision-free sidebands. 

It is worth emphasizing that by leaving the sweet spot, it is possible to use monochromatic modulation to attain most (but not all) of the $\bar{f}$ band available at bichromatic sweet spots. However, doing so comes at a significant cost to both $T_2^*$ (see Fig.~\ref{fig:coherence}) and gate stability over time. The advantage of bichromatic modulation is not simply the attainability of these $\bar{f}$ values, but rather the attainability of these $\bar{f}$ values at operating points protected from slow flux noise. 

\begin{figure}
    \includegraphics[width=\columnwidth]{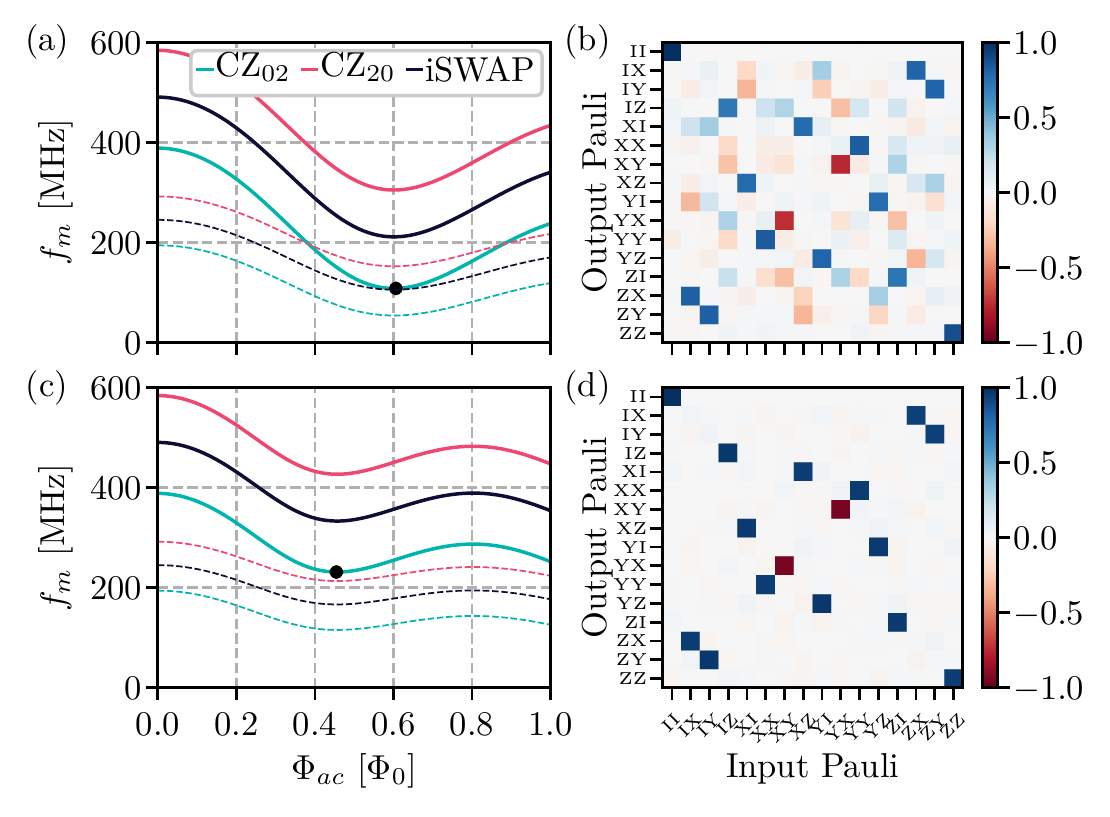}
    \caption{(a) Monochromatic modulation frequencies for activating native entangling gates between qubits 1 and 2 with the $k=-2$ (solid lines) and $k=-4$ (dashed lines) sidebands at different modulation amplitudes. A collision is found between the $k=-2$ $\mathrm{CZ}_{02}$ and $k=-4$ iSWAP at the sweet spot. (b) Measured process map for the optimized monochromatic $\mathrm{CZ}_{02}$ pulse, generated via process tomography. Strong XY-like terms are introducing coherent error. (c) Resonance conditions for the same edge, but under bichromatic modulation with $\alpha/2\pi = 0.085$ and $\theta/2\pi = -0.06$. The corresponding shift in $\bar{f}$ resolves the gate collision. (d) Measured process map for the optimized bichromatic $\mathrm{CZ}_{02}$ pulse, revealing a much cleaner approximation of the ideal CZ operator.} 
    \label{fig:collisions}
\end{figure}

Here we focus on qubits 1 and 2 (ref.~Table~\ref{qubit_table}), for which a sideband collision at the sweet spot limits fidelity in the monochromatic case. A native CZ interaction takes place when an $f_{01}$ sideband is placed on resonance with a capacitively-coupled neighbor's $f_{12}$ transition frequency (or vice versa), allowing $\ket{11} \leftrightarrow \ket{02}$ swapping (or $\ket{11} \leftrightarrow \ket{20}$ swapping). Similarly, a native iSWAP interaction takes place when an $f_{01}$ sideband is placed on resonance with a neighbor's $f_{01}$, allowing $\ket{10} \leftrightarrow \ket{01}$ swapping~\cite{Nico:2018}. For the pair in question, the various resonance conditions of the $k=-2$ and $k=-4$ sidebands produced by monochromatic modulation are shown in Fig.~\ref{fig:collisions}(a). At the sweet spot ($\Phi_{\ac} \approx 0.6\Phi_0$), the $k=-2$ $\text{CZ}_{02}$ interaction collides with the $k=-4$ iSWAP. While the $k=-2$ sideband has significantly more weight and the $\ket{11} \leftrightarrow \ket{02}$ transfer will happen more quickly as a result, cutting off the interaction at the ideal time for CZ still produces an operation that involves a strong XY-like swapping (see the process map in Fig.~\ref{fig:collisions}(b)~\cite{Chuang:2009,Nielsen:2002,Ryan_2015}). This CZ operation has a fidelity of $90.02 \pm 2.60\%$ as measured by interleaved randomized benchmarking~\cite{Knill_2008, Magesan_2012}.

However, a change in $\bar{f}$ can resolve this problem. If $\bar{f}$ is shifted by a value $\Delta$, then the modulation frequency required to activate $k=-2$ interactions moves by $\Delta/2$ while the frequency required for $k=-4$ interactions only moves by $\Delta/4$, according to Eq.~\eqref{sideband_freqs}. Using bichromatic modulation of $\alpha/2\pi = 0.085$ and $\theta/2\pi = -0.06$, the sweet spot modulation amplitude is reduced by about $25\%$ and the corresponding $\bar{f}$ is shifted up by $\Delta \approx 234\,\mathrm{MHz}$. The resulting collision-free set of resonance conditions is shown in Fig.~\ref{fig:collisions}(c), along with a much cleaner process map for the calibrated CZ gate in Fig.~\ref{fig:collisions}(d). Gate error is reduced by a factor 5 from the monochromatic case, reaching an interleaved randomized benchmarking fidelity of $98.06 \pm 0.28 \%$. 

Furthermore, this particular set of bichromatic parameters is not unique in their ability to dodge the gate collision that limits monochromatic fidelity. As long as no new collisions are introduced or gate time is not significantly lengthened at a specific operating point, $\bar{f}$ can be placed anywhere across the attainable bichromatic sweet spot band. By simulating the detuning expected for modulation of various $\alpha$ and $\theta$, gates can be found at specific, desirable values of $\bar{f}$ or modulation frequencies $f_m$. 

\begin{figure}
    \includegraphics[width=\columnwidth]{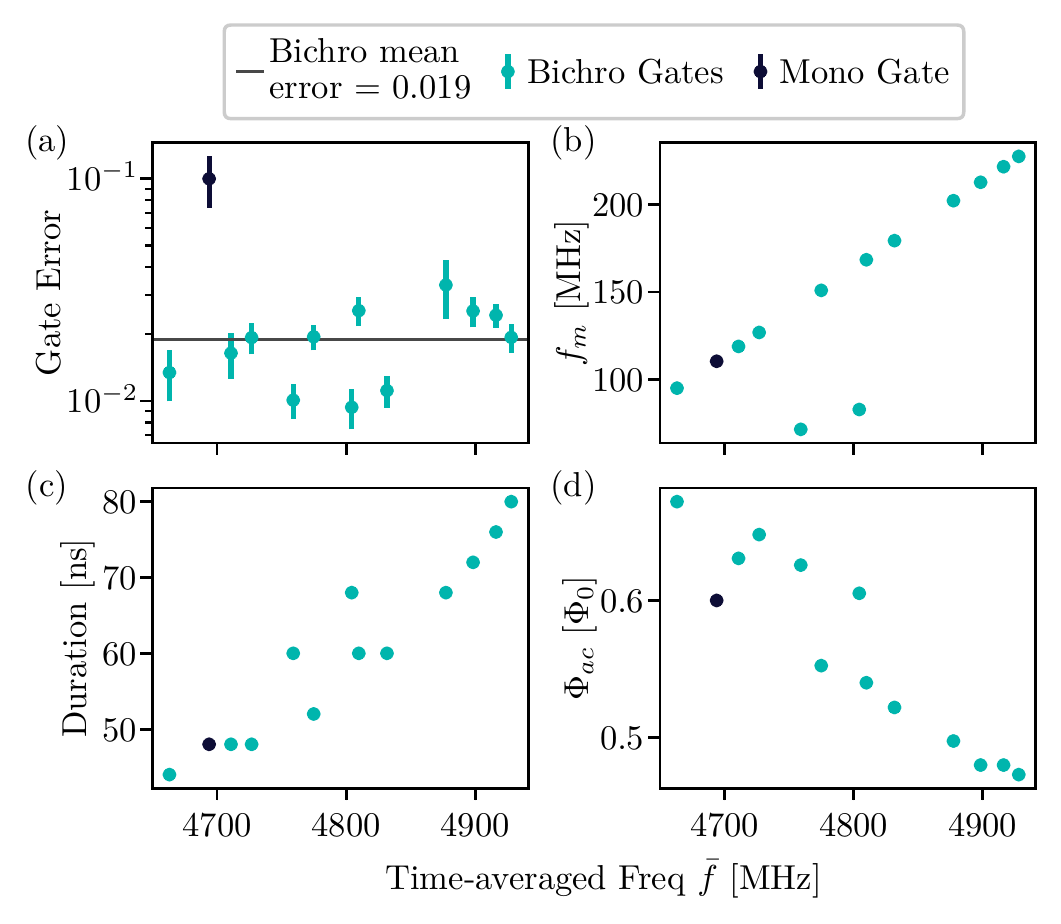}
    \caption{(a) The gate error for 12 bichromatic CZs between qubit 1 and qubit 2, measured via interleaved randomized benchmarking and plotted against time-averaged frequency of qubit 1 during modulation. The gates include both $k=-2$ and $k=-4$ sideband interactions, all operated at dynamical sweet spots. The $k=-2$ gate at the single monochromatic sweet spot is included for comparison. The monochromatic $k=-4$ gate was not brought up due to the required $f_m$ being too low for the filtration scheme in use. (b) Modulation frequencies $f_m$ for each operating point. The distinct linear sets correspond to the two sidebands being used to activate the gates. (c) Gate times for the full set of operating points. The native CZ interaction involves two full $\ket{11} \leftrightarrow \ket{02}$ swaps, the rate of which is determined by the weight of the activating sideband. (d) Modulation amplitudes $\Phi_{\ac}$, all of which satisfy the dynamical sweet spot condition $\partial \bar{f}/\partial \Phi_{\ac} = 0$ for each set of modulation parameters.}    
    \label{fig:bichro-gates}
\end{figure}

A set of CZ gates operated across a range of bichromatic operating points is shown in  Fig.~\ref{fig:bichro-gates}. All of these gates are operated at sweet spots, and the corresponding average frequencies span $264\,\mathrm{MHz}$. Via interleaved randomized benchmarking, this representative sampling of gates is found to reach a mean fidelity of just over 98\%, with individual gates as high as $99.06 \pm 0.19\%$. In addition to error rates, Fig.~\ref{fig:bichro-gates} shows the modulation frequencies, durations, and modulation amplitudes for each gate. 

Again, it is worth noting that these fidelities are certainly not achievable at every sweet spot in the available continuum, as there are many such operating points that suffer from collisions just like the monochromatic gate, or that result in poor weighting of the activating sideband.  The set of gates in Fig.~\ref{fig:bichro-gates} is meant only to communicate a lower bound on the density with which gates of at least these fidelities can be brought up. This flexibility in operating point represents a significant advantage in robustness over the single monochromatic sweet spot, and will only become more useful as higher lattice connectivity introduces more frequency crowding.

\section{Sideband Engineering for Faster Gates}
\label{sidebands}

Bichromatic control of parametric gates also provides an effective tool for optimizing sideband weights and reducing gate times. 
During flux modulation, the static coupling between the modulated qubit and each of its neighbors is renormalized across all generated sidebands according to their weights. This results in an effective coupling $g_{\text{eff}} = g\varepsilon_k$ for iSWAP interactions and $g_{\text{eff}} = \sqrt{2}g\varepsilon_k$ for CZ interactions, where $g$ is the bare coupling and $\varepsilon_k$ is the weight of the $k^{th}$ sideband~\cite{Nico:2018}. These weights are given by the coefficient of each harmonic in the Fourier expansion of the modulated qubit charge operator (see Appendix~\ref{theory}). The $g_{\text{eff}}$ for a monochromatic gate at the sweet spot is determined solely by the sideband used, and cannot be dynamically tuned because all modulation parameters are fixed. However, different $\alpha$ and $\theta$ alter the time-dependence of the qubit frequency under bichromatic modulation and can be used to tune the distribution of weight, even concentrating it into specific sidebands~\cite{Nico:bichro}. 

\begin{figure}
    \includegraphics[width=\columnwidth]{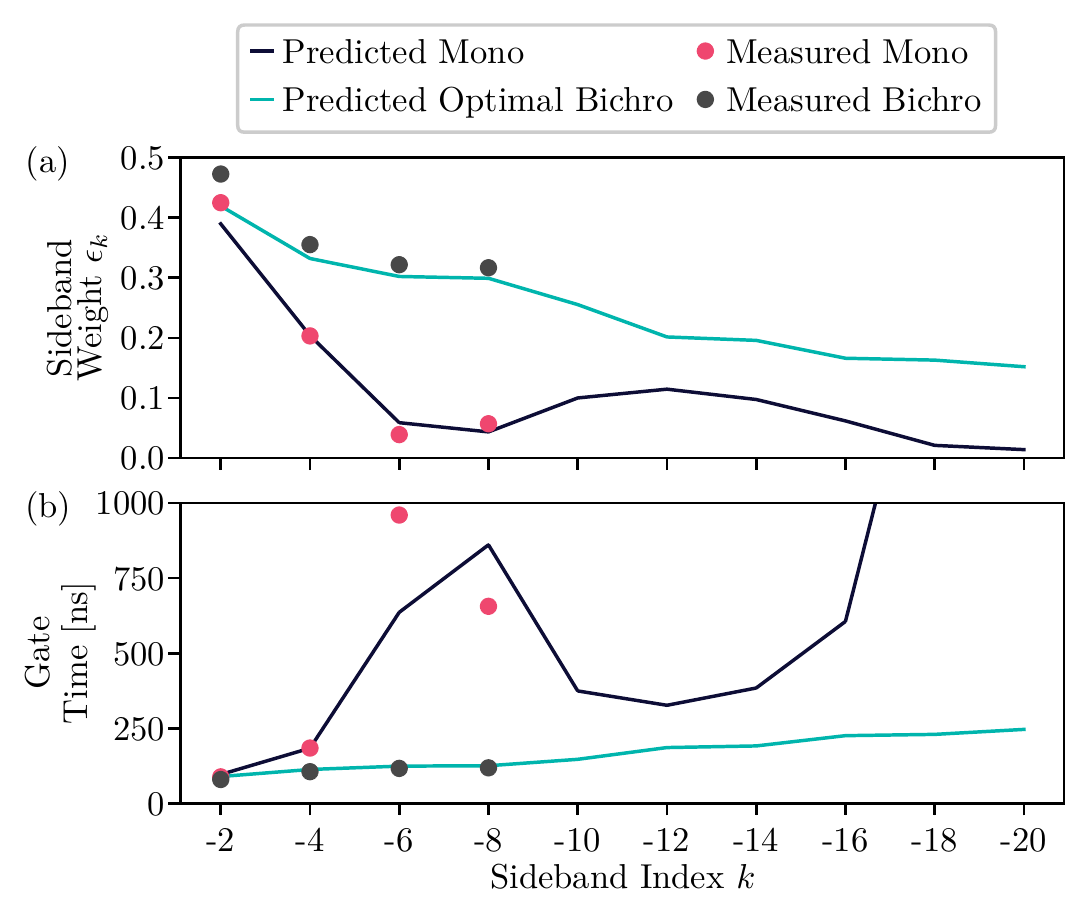}
    \caption{(a) Predicted and measured sideband weights achievable with monochromatic and bichromatic modulation of qubit 3. All weights correspond to sweet spot flux operating points for activating a $\text{CZ}_{02}$ interaction with qubit 4. The measured weights are extracted from the measured gate times. (b) The predicted and measured gate times associated with the sidebands described in panel (a). In general, bichromatic operating points can be optimized to concentrate weight into the gate-activating sideband, dramatically reducing obtainable gate times at higher sideband indices.}  
    \label{fig:sideband_weights}
\end{figure}

Fig.~\ref{fig:sideband_weights}(a) shows the modeled difference in achievable sideband weights using monochromatic versus bichromatic modulation to activate sweet spot $\text{CZ}_{02}$ interactions between qubit 3 (modulated) and qubit 4 (ref.~Table~\ref{qubit_table}). Each bichromatic operating point has been optimized to maximize the weight of the activating sideband. Only even-index sidebands are shown because qubit 3 is biased at the maximum of its tunability band for the scope of this study, and the symmetry of modulation around this parking frequency reduces the odd Fourier coefficients to 0. The trend in monochromatic weights is typical in that they oscillate downward with increasing sideband index, eventually vanishing. Comparatively, the optimized weights produced by specific bichromatic pulse parameters are similar for $k=-2$ but drop off less quickly with index. This leads to significant reductions in predicted gate times, which are inversely proportional to sideband weights and are displayed in Fig.~\ref{fig:sideband_weights}(b).

For monochromatic and bichromatic gates at each of the predicted operating points up through the $-8^{th}$ sideband, measured gate times and extracted sideband weights are plotted over the model results in Fig.~\ref{fig:sideband_weights}. We find good agreement for the sidebands with weights above 0.1, measuring increasingly long monochromatic CZs and relatively constant-duration bichromatic CZs with increasing sideband index. For the more weakly-weighted monochromatic $k=-6$ and $k=-8$ sidebands, small errors in predicted weight create larger discrepancies in gate times. That being said, these gates are found to be significantly longer than their bichromatic counterparts. 

\begin{figure}
    \includegraphics[width=\columnwidth]{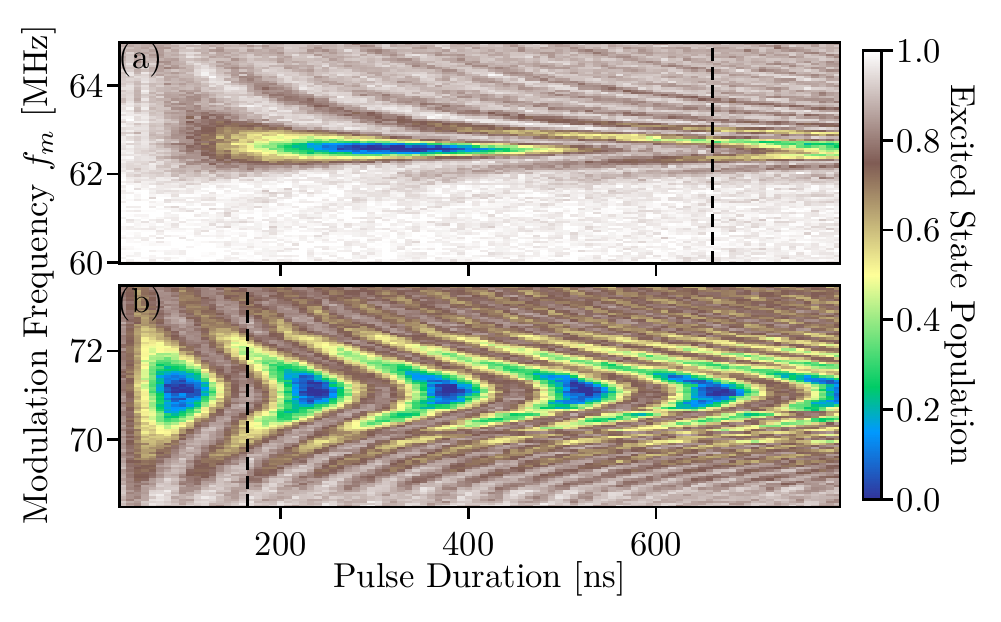}
    \caption{Excited state population of qubit 4, undergoing $\ket{1} \leftrightarrow \ket{0}$ transfer during a CZ interaction mediated by the $k=-8$ sideband of qubit 3's $f_{12}$ transition frequency. Population exchange at each resonance condition is visualized by sweeping pulse duration and fundamental modulation frequency for constant modulation amplitude at the respective sweet spot. Panel (a) features the monochromatic instance of this gate, while panel (b) features a bichromatic operating point selected for increased sideband weight $\varepsilon_{-8}$ (not the same operating point featured in Fig.~\ref{fig:sideband_weights}, which has lower fidelity due to a collision). The frequency axes of both panels have the same scale, highlighting the relative resonance widths that scale inversely to coupling strength. The black dotted lines correspond to fit durations of a full CZ interaction, showing a bichromatic speed-up by almost exactly a factor of 4 after pulse risetimes are included in total duration. The bichromatic gate has an interleaved randomized benchmarking fidelity of $98.85 \pm 0.50 \%$, compared to the $92.74 \pm 2.15 \%$ fidelity of the monochromatic gate.}  
    \label{fig:chevrons}
\end{figure}

The population exchange of qubit 4 during two $k=-8$ $\text{CZ}_{02}$ interactions with qubit 3 is shown in Fig.~\ref{fig:chevrons}, with the monochromatic sweet spot operating point in panel (a) and a faster bichromatic sweet spot operating point in panel (b). While the activating fundamental modulation frequencies are only about 8.5 MHz apart, the difference in $\bar{f}$ is 8 times that due to the sideband being used. The bichromatic operating point provides a factor of 4 speed-up as well as a corresponding boost in interleaved randomized benchmarking fidelity, from $92.74 \pm 2.15 \%$ for the monochromatic gate to $98.85 \pm 0.50 \%$ for the bichromatic. Measurements of coherence during modulation show the fidelity of both gates to be coherence-limited within uncertainty~\cite{Nico:2018}. 

While it would be rare, at least in the context of Hamiltonians similar to the one under study, to produce a pair of qubits for which an optimized bichromatic $k=-8$ CZ is faster than any $k=-2$ alternative, the ability to significantly increase coupling to higher-order sidebands is not without benefit. Upper limits on $f_m$ due to AWG bandwidth mean that using higher order sidebands can both unlock additional $\bar{f}$ flexibility as well as activate interactions that would otherwise be out of band. Bringing those gates into the realm of viability by reducing their duration is another impactful use-case for bichromatic modulation.  

\section{Conclusion}

We have experimentally verified that bichromatic flux modulation can be used to extend the range of time-averaged frequencies accessible to tunable transmons at operating points protected from slow flux noise. We have demonstrated the usefulness of this freedom for reducing two-qubit parametric gate error, both by dodging unwanted collisions and reducing gate speed with improved sideband weights. 

Extension of bichromatic flux control to access the full XY and CPHASE gate families~\cite{Deanna_2020} is the subject of upcoming work. Another promising application is parametric-resonance gates, which have recently been implemented on a tunable-coupler architecture~\cite{Sete:2021}. To perform these gates, the average frequency of a modulated qubit is brought directly onto resonance with a neighbor. With the frequency of modulation no longer relevant for activating the gate, it can be optimized independently to avoid collisions. Moreover, the weight of the activating $k=0$ sideband can reach values close to unity, allowing fast gates at the bare coupling rate. Bichromatic control would allow these gates to be implemented at dynamical sweet spots, presenting the opportunity to increase their fidelity substantially~\cite{Nico:bichro}.

In summary, the flexibility provided by this novel control technique represents an important step forward in terms of robustness to experimental imperfection as well as scalability of flux-tunable qubit architectures. As such, bichromatic flux modulation is shown to be a promising method for achieving controlled entanglement on superconducting processors. 

\section*{Acknowledgements}

We thank Prasahnt Sivarajah and Deanna Abrams for their initial exploration of calibrating bichromatic pulses. We thank Gregory Stiehl and Alexander Hill for helpful discussions and for critically reading the manuscript. We thank Ben Scharmann for his help in designing the device Hamiltonians under study. Finally we thank all of our collaborators at Rigetti Computing, without whose effort the computing stack used in this study would not exist.

The experimental results presented here are based upon work supported by the Defense Advanced Research Projects Agency (DARPA) under agreement No.~HR00112090058.

\section*{Contributions}

J.A.V. drafted the manuscript. S.C. and J.A.V. performed the initial validation of theory. J.A.V. implemented the measurement tools and performed the experiments. N.D. developed the theory describing bichromatic flux control of tunable transmons and created simulation tools used in the study. G.J. organized the construction of the control systems in use, and provided critical knowledge and support throughout the study. N.D. organized the experimental effort.

\appendix

\section{Pulse Calibration}
\label{pulse_calibration}

In achieving the results discussed in this paper, the bichromatic pulses defined in Sec.~\ref{sweet_spots} are produced by a single FPGA-based sequencer for simplicity. The modulation waveform containing both frequencies of the bichromatic pulse is precomputed as a single set of digital values which are fed into a DAC. This scheme is sufficient for the experiments performed, but requires some additional calibration, described here.

It is normally convenient, for the sake of tracking phase, to define control pulses relative to local rotating frames that oscillate at some characteristic frequency of the system. Doing so imparts a global frame phase on those pulses, which is usually irrelevant but introduces complications for bichromatic pulses as we have constructed them, with two frequencies in the baseband waveform. Specifically, the effective value of $\theta$ is shifted by $1-p$ times the applied phase.

To understand this, consider a simplified version of the AC pulse in Eq.~\eqref{general_waveform} with a global phase shift $\beta$: \begin{equation}
    \begin{split}
        \Phi(t) = &\cos(\alpha)\cos\big{(}2{\pi}f_mt+\beta\big{)}\\+ \text{ }&\sin(\alpha)\cos\big{(}2{\pi}pf_mt + \theta + \beta\big{)}.
    \end{split}
    \label{simple_shifted_waveform}
\end{equation}
To reorganize into the conventional format of a single relative phase on the second tone, we can advance the clock by $\beta/(2{\pi}f_m)$, the amount of time it takes the oscillation of the first tone to cycle through $\beta$. With $t' = t + \frac{\beta}{2{\pi}f_m}$, we can re-express Eq.~\eqref{simple_shifted_waveform} as 
\begin{equation}
    \begin{split}
        \Phi(t') = &\cos(\alpha)\cos\big{(}2{\pi}f_mt'\big{)}\\+ \text{ }&\sin(\alpha)\cos\big{(}2{\pi}pf_mt' + \theta + (1-p)\beta\big{)}.
    \end{split}
    \label{simple_shifted_waveform_t'}
\end{equation}
In this form, it becomes clear that the $\beta$ phase shift has produced a new effective relative phase between the tones, $\theta' = \theta + (1-p)\beta$. In terms of pre-compensation, the used $\theta$ should be $(p-1)\beta$ greater than the desired $\theta$.

Another way to understand this is that every phase shift $\beta$ of the $f_m$ tone should correspond to a phase shift $p\beta$ of the $pf_m$ tone if the relative phase between the tones is to be preserved. When both tones are shifted by $\beta$, there is a $(p-1)\beta$ difference to make up.

\begin{figure}
    \includegraphics[width=\columnwidth]{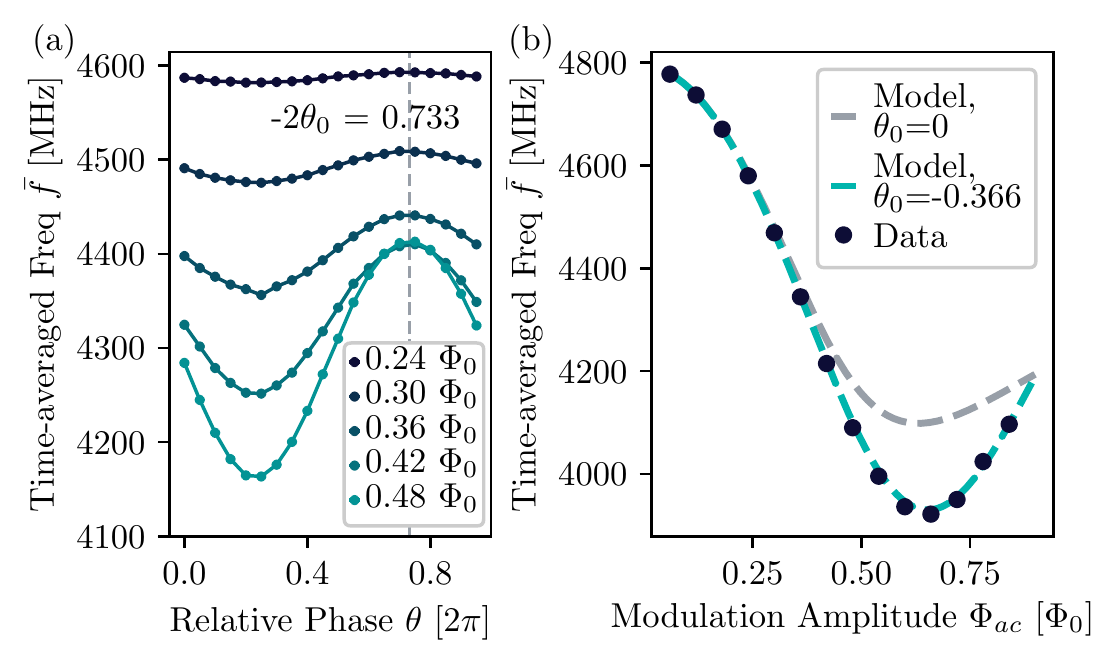}
    \caption{(a) Measurement of $\theta_0$ for qubit 3. For bichromatic modulation of constant $p$, $\alpha$, and amplitude $\Phi_{\ac}$, time-averaged qubit frequency $\bar{f}$ is sinusoidal in $\theta+(1-p)\theta_0$. The phase offset of the fitted $\bar{f}$ oscillation in $\theta$ can thus be mapped to $\theta_0$. While one $\Phi_{\ac}$ is sufficient to measure this offset, the fit value here is the median of offsets measured at five different modulation amplitudes. (b) For the same qubit, measured detuning under bichromatic modulation is compared against model predictions that do and do no take into account the measured $\theta_0$. The models agree at low modulation amplitudes, which can also be seen in panel (a) in the form of smaller fluctuations in $\bar{f}$ at smaller $\Phi_{\ac}$. Close to the sweet spot however, $\theta_0$ corresponds to a predicted detuning difference of about $183\,\mathrm{MHz}$. The data matches the model that incorporates the measured $\theta_0$ value, illustrating the importance of this calibration for experimental realization of predicted behavior.}    
    \label{fig:theta_0}
\end{figure}

In the case of a global phase coming from a rotating frame on which a bichromatic pulse is defined, the necessary $\theta$ pre-compensation is inconvenient because the applied shift will change depending on when a pulse is played in a program. As a result, we forego a rotating frame completely and instead define the bichromatic modulation tones relative to the lab frame. This eliminates the overhead involved in redefining pulses based on a tracked frame phase throughout a program.

\begin{figure}
    \includegraphics[width=\columnwidth]{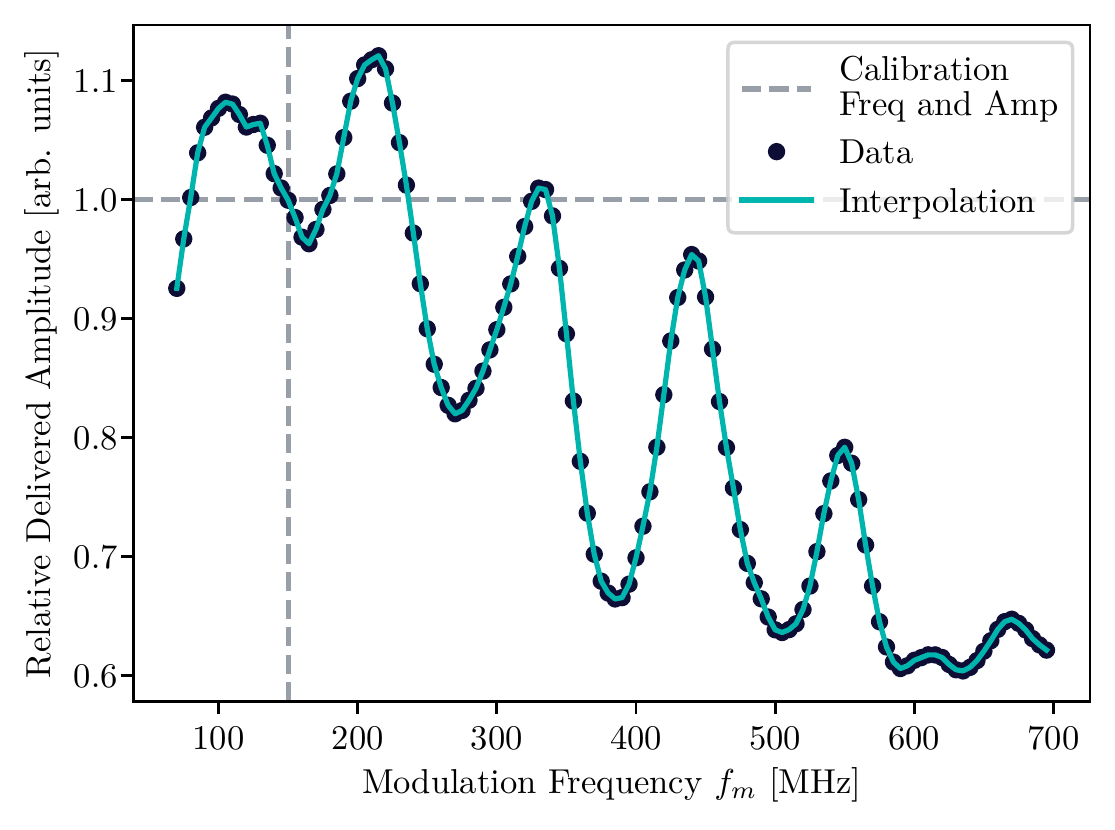}
    \caption{An example amplitude transfer function of a flux transmission line, for qubit 3 in this case (ref.~Table~\ref{qubit_table}). At each frequency, the effective modulation amplitude seen by the qubit is determined by measuring the detuning of the qubit frequency. The strong frequency-dependence of the transfer function can then be compensated for by scaling flux pulses according to their frequency and its measured relative transmission.}    
    \label{fig:transfer_function}
\end{figure}

However, even with no frame phase being applied, the relative phase between the FPGA clock and local oscillator (LO) within each waveform generator introduces its own global phase shift on the bichromatic waveform. Let this phase be denoted as $\theta_0$. Unlike the frame phase, it is possible to keep $\theta_0$ constant by ensuring all pulse programs start on integer multiples of both the FPGA clock cycle and the LO period. With these conditions met, $\theta_0$ becomes a constant phase that needs to be calibrated once per AWG in order to accurately produce specific values of $\theta$ in applied flux of the following form:

\begin{equation}
    \begin{split}
        \Phi(t) = &\Phi_{\dc} +\Phi_{\ac}u(t)\Big{[}\cos(\alpha)\cos\big{(}2{\pi}f_mt\big{)} +\\ \text{}&\sin(\alpha)\cos\big{(}2{\pi}pf_mt + \theta +(1-p)\theta_0\big{)}\Big{]}.
    \end{split}
    \label{theta_0_waveform}
\end{equation}

The constant phase offset can be measured by sweeping $\theta$ for modulation of constant $\Phi_{\ac}$ and measuring the induced detuning, which will have sinusoidal dependence on $\theta$ with an offset of $(1-p)\theta_0$ (See Appendix~\ref{theory}). Average qubit detuning during modulation is measured with a Ramsey-style experiment in which each flux pulse is continually delivered during the delay times between $\RX(\pi/2)$ rotations. Fig.~\ref{fig:theta_0}(a) shows the results of one such calibration measurement for qubit 3 (ref.~Table~\ref{qubit_table}), while Fig.~\ref{fig:theta_0}(b) communicates the importance of accurately measuring and accounting for $\theta_0$ when attempting to reconcile model and data. 

We have verified these calibrations to be stable in time on the order of at least weeks in our setup. However, it is worth noting that if restarting an AWG randomizes the clock-oscillator phase, it would be necessary to remeasure~$\theta_0$.

Regarding the mixing angle $\alpha$, achieving a desired value requires compensation for any frequency-dependent attenuation present in the flux line. The calibration of this transfer function is performed with the same flux-Ramsey procedure described above, where the different detunings caused by monochromatic modulation of different frequencies are used to determine corrective scaling factors. An example transfer function can be found in Fig.~\ref{fig:transfer_function}. In the construction of bichromatic pulses, the amplitude of each tone must be scaled appropriately for its frequency before the tones are digitally added together.

\begin{figure}
    \includegraphics[width=\columnwidth]{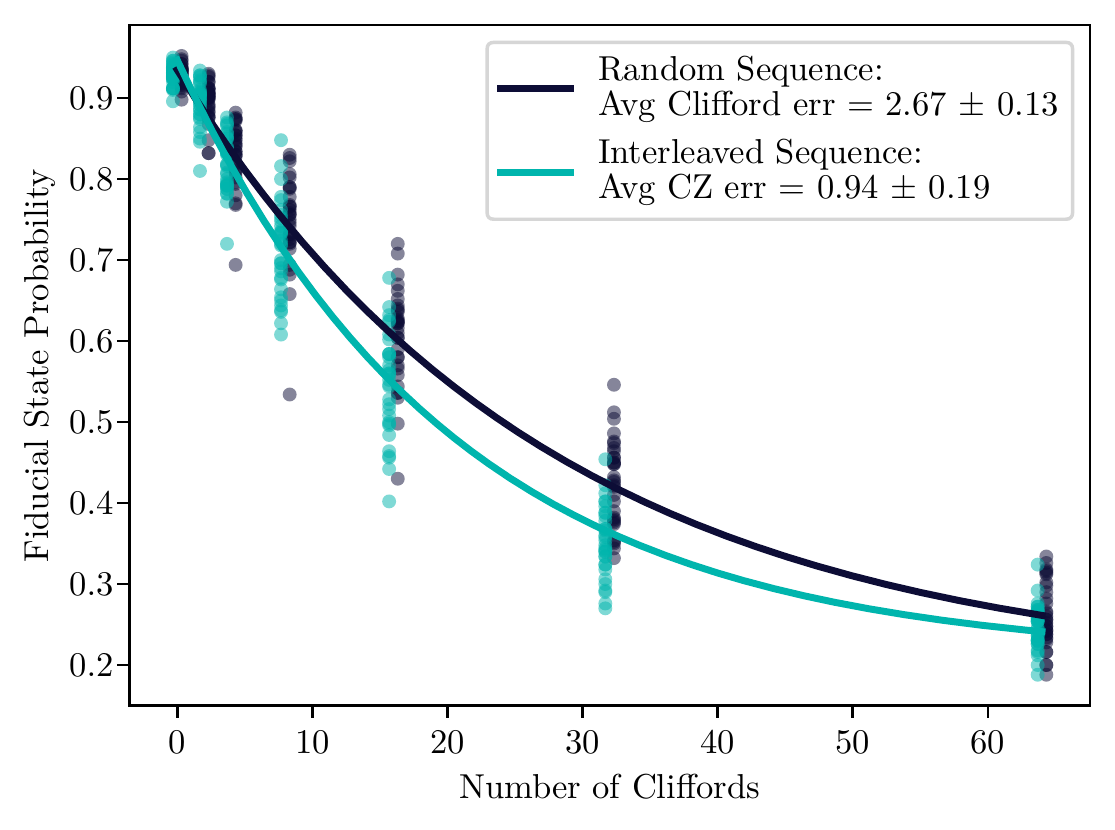}
    \caption{An example of the interleaved randomized benchmarking data used to produce the fidelities in Table~\ref{gate_table}. This particular data corresponds to the gate during which qubit 1 average frequency $\bar{f} = 4804\,\mathrm{MHz}$. 32 trials are run for each sequence length of 0, 2, 4, 8, 16, 32, and 64 cliffords. Individual data points correspond to the average probability of reading out the fiducial state over 500 shots per trial.} 
    \label{fig:rb}
\end{figure}

\begin{table*}[t]
\begin{ruledtabular}
\begin{tabular}{c c c c c c c c}
\shortstack{Frequency \\ Multiplier $p$} & \shortstack{Mixing \\ Angle $\alpha$ [2$\pi$]} & \shortstack{Relative \\ Phase $\theta$ [2$\pi$]} & \shortstack{Time-averaged \\ Frequency \\ $\bar{f}$ [MHz]} & \shortstack{Modulation \\ Frequency \\$f_m$ [MHz]} & \shortstack{Interaction \\ Time [ns]} & \shortstack{Modulation \\ Amplitude \\$\Phi_{\ac}$ [$\Phi_0$]} & \shortstack{Gate \\ Fidelity [\%]} \\
\hline
3 & 0.005 & 0.570 & 4663 & 95.04 & 44 & 0.67 & 98.66 $\pm$ 0.34 \\
0 & 0.000 & 0.000 & 4694 & 110.41 & 48 & 0.60 & 90.02 $\pm$ 2.60 \\
3 & 0.005 & 0.045 & 4711 & 118.88 & 48 & 0.63 & 98.36 $\pm$ 0.38 \\
3 & 0.015 & 0.210 & 4727 & 126.87 & 48 & 0.65 & 98.07 $\pm$ 0.31 \\
3 & 0.015 & 0.200 & 4759 & 71.50 & 60 & 0.63 & 98.99 $\pm$ 0.18 \\
3 & 0.020 & -0.050 & 4775 & 150.93 & 52 & 0.55 & 98.06 $\pm$ 0.26 \\
3 & 0.025 & 0.150 & 4804 & 82.84 & 68 & 0.61 & 99.06 $\pm$ 0.19 \\
3 & 0.035 & -0.100 & 4810 & 168.42 & 60 & 0.54 & 97.45 $\pm$ 0.38 \\
3 & 0.035 & 0.000 & 4832 & 179.33 & 60 & 0.52 & 98.89 $\pm$ 0.18 \\
3 & 0.050 & -0.045 & 4877 & 202.17 & 68 & 0.50 & 96.68 $\pm$ 0.99 \\
3 & 0.060 & -0.032 & 4898 & 212.70 & 72 & 0.48 & 97.46 $\pm$ 0.39 \\
3 & 0.075 & -0.056 & 4916 & 221.59 & 76 & 0.48 & 97.57 $\pm$ 0.30 \\
3 & 0.085 & -0.060 & 4928 & 227.49 & 80 & 0.47 & 98.07 $\pm$ 0.28 \\
\end{tabular}
\end{ruledtabular}
\caption{Information about the CZ operating points and gates presented in Sec.~\ref{gates}, listed in order of qubit 1 $\bar{f}$. All of these operating points represent dynamical sweet spots with first-order protection against $1/f$ flux noise.}
\label{gate_table}
\end{table*}

\section{Transmon under bichromatic modulation}
\label{theory}

The properties of a tunable transmon are controlled with magnetic flux applied on a SQUID loop by changing the effective Josephson energy
\begin{align}
E_{J,\mathrm{eff}}=\sqrt{E_{J,1}^2+E_{J,2}^2+2E_{J,1}E_{J,2}\cos(\phi)}.
\label{EJeff}
\end{align}
Here, $\phi=2\pi\Phi/\Phi_0$ and $E_{J,1}$, $E_{J,2}$ are the Josephson energies of the SQUID loop.
The transition frequencies are $\Phi_0$-periodic, symmetric around zero flux bias, and can be expressed as the Fourier series
\begin{align}
f(\phi) &= \sum_{n=0}^\infty F_{n} \cos(n\phi).
\end{align}
Under a bichromatic modulation 
$\Phi(t) = \Phi_\dc + \Phi_{\ac1}\cos(2\pi f_mt+\theta_1) + \Phi_{\ac,p}\cos(2\pi pf_mt+\theta_p)$, 
the time-averaged transmon frequency is equal to
\begin{align}
\bar{f} =& \sum_{m=0}^\infty \nu_m \cos(m\theta), \label{favgFourier}\\
\nu_m =& \sum_{n=0}^\infty F_{n} \cos[n\phi_\dc+(p+1)m\tfrac{\pi}{2}] \nonumber\\ 
& \times (2-\delta_{m,0}) \mathrm{J}_{pm}(n\phi_{\ac,1})\mathrm{J}_{m}(n\phi_{\ac,p}), \nonumber
\end{align}
which is $2\pi$-periodic in $\theta = \theta_p-p\theta_1$.
Given this symmetry, we choose the convention $\theta_1=0$ in defining bichromatic pulses throughout the study.

In the transmon basis, the qubit-qubit coupling is flux-dependent.
Under modulation, the effective coupling activated by the $k^\text{th}$ sideband is $g_k = g \varepsilon_k$, renormalized from the zero-flux bare coupling $g$ by the sideband weight $\varepsilon_k$. These weights are obtained from
\begin{align}
g_k = \frac{1}{T_m}\int_0^{T_m}\mathrm{d}t\, g(\Phi(t)) e^{i2\pi\int_0^t\mathrm{d}t'f(\Phi(t'))}e^{-i2\pi(\bar{f}+kf_m)t},
\end{align}
where $T_m=1/f_m$ is the modulation period.

Throughout the experiments described in this study, $\Phi_\dc=0$ and $p$ is odd. The bichromatic modulation is then $T_m/2$-antiperiodic, and because the effective Josephson energy Eq.~\eqref{EJeff} is an even function of the flux bias, the transmon parameters oscillate at twice the modulation frequency.
As a consequence, only the even sidebands are generated.

Moreover, for $p=3$, the dependence of the average frequency on the DC bias in Eq.~\eqref{favgFourier} simplifies to $\cos(n\phi_\dc)$. Its derivative vanishes at $\Phi_\dc=0$, thereby protecting the transmon from slow additive flux noise during modulation.

\section{Gate Information}
\label{gate_info}

For reference, included in Table~\ref{gate_table} are details regarding the bichromatic operating points used to perform the gates presented in Sec.~\ref{gates}. An example instance of interleaved randomized benchmarking for one of these gates can be seen in Fig.~\ref{fig:rb}, along with implementation details in the caption~\cite{Knill_2008, Magesan_2012}. All fidelities cited in this manuscript were measured using this same protocol.


\begin{thebibliography}{67}%
\makeatletter
\providecommand \@ifxundefined [1]{%
 \@ifx{#1\undefined}
}%
\providecommand \@ifnum [1]{%
 \ifnum #1\expandafter \@firstoftwo
 \else \expandafter \@secondoftwo
 \fi
}%
\providecommand \@ifx [1]{%
 \ifx #1\expandafter \@firstoftwo
 \else \expandafter \@secondoftwo
 \fi
}%
\providecommand \natexlab [1]{#1}%
\providecommand \enquote  [1]{``#1''}%
\providecommand \bibnamefont  [1]{#1}%
\providecommand \bibfnamefont [1]{#1}%
\providecommand \citenamefont [1]{#1}%
\providecommand \href@noop [0]{\@secondoftwo}%
\providecommand \href [0]{\begingroup \@sanitize@url \@href}%
\providecommand \@href[1]{\@@startlink{#1}\@@href}%
\providecommand \@@href[1]{\endgroup#1\@@endlink}%
\providecommand \@sanitize@url [0]{\catcode `\\12\catcode `\$12\catcode
  `\&12\catcode `\#12\catcode `\^12\catcode `\_12\catcode `\%12\relax}%
\providecommand \@@startlink[1]{}%
\providecommand \@@endlink[0]{}%
\providecommand \url  [0]{\begingroup\@sanitize@url \@url }%
\providecommand \@url [1]{\endgroup\@href {#1}{\urlprefix }}%
\providecommand \urlprefix  [0]{URL }%
\providecommand \Eprint [0]{\href }%
\providecommand \doibase [0]{https://doi.org/}%
\providecommand \selectlanguage [0]{\@gobble}%
\providecommand \bibinfo  [0]{\@secondoftwo}%
\providecommand \bibfield  [0]{\@secondoftwo}%
\providecommand \translation [1]{[#1]}%
\providecommand \BibitemOpen [0]{}%
\providecommand \bibitemStop [0]{}%
\providecommand \bibitemNoStop [0]{.\EOS\space}%
\providecommand \EOS [0]{\spacefactor3000\relax}%
\providecommand \BibitemShut  [1]{\csname bibitem#1\endcsname}%
\let\auto@bib@innerbib\@empty
\bibitem [{\citenamefont {Martinis}\ \emph {et~al.}(2005)\citenamefont
  {Martinis}, \citenamefont {Cooper}, \citenamefont {McDermott}, \citenamefont
  {Steffen}, \citenamefont {Ansmann}, \citenamefont {Osborn}, \citenamefont
  {Cicak}, \citenamefont {Oh}, \citenamefont {Pappas}, \citenamefont
  {Simmonds},\ and\ \citenamefont {Yu}}]{Martinis_2005}%
  \BibitemOpen
  \bibfield  {author} {\bibinfo {author} {\bibfnamefont {J.~M.}\ \bibnamefont
  {Martinis}}, \bibinfo {author} {\bibfnamefont {K.~B.}\ \bibnamefont
  {Cooper}}, \bibinfo {author} {\bibfnamefont {R.}~\bibnamefont {McDermott}},
  \bibinfo {author} {\bibfnamefont {M.}~\bibnamefont {Steffen}}, \bibinfo
  {author} {\bibfnamefont {M.}~\bibnamefont {Ansmann}}, \bibinfo {author}
  {\bibfnamefont {K.~D.}\ \bibnamefont {Osborn}}, \bibinfo {author}
  {\bibfnamefont {K.}~\bibnamefont {Cicak}}, \bibinfo {author} {\bibfnamefont
  {S.}~\bibnamefont {Oh}}, \bibinfo {author} {\bibfnamefont {D.~P.}\
  \bibnamefont {Pappas}}, \bibinfo {author} {\bibfnamefont {R.~W.}\
  \bibnamefont {Simmonds}},\ and\ \bibinfo {author} {\bibfnamefont {C.~C.}\
  \bibnamefont {Yu}},\ }Decoherence in josephson qubits from dielectric loss,\
  \href {https://doi.org/10.1103/PhysRevLett.95.210503} {\bibfield  {journal}
  {\bibinfo  {journal} {Phys. Rev. Lett.}\ }\textbf {\bibinfo {volume} {95}},\
  \bibinfo {pages} {210503} (\bibinfo {year} {2005})}\BibitemShut {NoStop}%
\bibitem [{\citenamefont {M\"uller}\ \emph {et~al.}(2015)\citenamefont
  {M\"uller}, \citenamefont {Lisenfeld}, \citenamefont {Shnirman},\ and\
  \citenamefont {Poletto}}]{Muller_2015}%
  \BibitemOpen
  \bibfield  {author} {\bibinfo {author} {\bibfnamefont {C.}~\bibnamefont
  {M\"uller}}, \bibinfo {author} {\bibfnamefont {J.}~\bibnamefont {Lisenfeld}},
  \bibinfo {author} {\bibfnamefont {A.}~\bibnamefont {Shnirman}},\ and\
  \bibinfo {author} {\bibfnamefont {S.}~\bibnamefont {Poletto}},\ }Interacting
  two-level defects as sources of fluctuating high-frequency noise in
  superconducting circuits,\ \href {https://doi.org/10.1103/PhysRevB.92.035442}
  {\bibfield  {journal} {\bibinfo  {journal} {Phys. Rev. B}\ }\textbf {\bibinfo
  {volume} {92}},\ \bibinfo {pages} {035442} (\bibinfo {year}
  {2015})}\BibitemShut {NoStop}%
\bibitem [{\citenamefont {Klimov}\ \emph {et~al.}(2018)\citenamefont {Klimov},
  \citenamefont {Kelly}, \citenamefont {Chen}, \citenamefont {Neeley},
  \citenamefont {Megrant}, \citenamefont {Burkett}, \citenamefont {Barends},
  \citenamefont {Arya}, \citenamefont {Chiaro}, \citenamefont {Chen},
  \citenamefont {Dunsworth}, \citenamefont {Fowler}, \citenamefont {Foxen},
  \citenamefont {Gidney}, \citenamefont {Giustina}, \citenamefont {Graff},
  \citenamefont {Huang}, \citenamefont {Jeffrey}, \citenamefont {Lucero},
  \citenamefont {Mutus}, \citenamefont {Naaman}, \citenamefont {Neill},
  \citenamefont {Quintana}, \citenamefont {Roushan}, \citenamefont {Sank},
  \citenamefont {Vainsencher}, \citenamefont {Wenner}, \citenamefont {White},
  \citenamefont {Boixo}, \citenamefont {Babbush}, \citenamefont {Smelyanskiy},
  \citenamefont {Neven},\ and\ \citenamefont {Martinis}}]{Klimov:2018}%
  \BibitemOpen
  \bibfield  {author} {\bibinfo {author} {\bibfnamefont {P.~V.}\ \bibnamefont
  {Klimov}}, \bibinfo {author} {\bibfnamefont {J.}~\bibnamefont {Kelly}},
  \bibinfo {author} {\bibfnamefont {Z.}~\bibnamefont {Chen}}, \bibinfo {author}
  {\bibfnamefont {M.}~\bibnamefont {Neeley}}, \bibinfo {author} {\bibfnamefont
  {A.}~\bibnamefont {Megrant}}, \bibinfo {author} {\bibfnamefont
  {B.}~\bibnamefont {Burkett}}, \bibinfo {author} {\bibfnamefont
  {R.}~\bibnamefont {Barends}}, \bibinfo {author} {\bibfnamefont
  {K.}~\bibnamefont {Arya}}, \bibinfo {author} {\bibfnamefont {B.}~\bibnamefont
  {Chiaro}}, \bibinfo {author} {\bibfnamefont {Y.}~\bibnamefont {Chen}},
  \bibinfo {author} {\bibfnamefont {A.}~\bibnamefont {Dunsworth}}, \bibinfo
  {author} {\bibfnamefont {A.}~\bibnamefont {Fowler}}, \bibinfo {author}
  {\bibfnamefont {B.}~\bibnamefont {Foxen}}, \bibinfo {author} {\bibfnamefont
  {C.}~\bibnamefont {Gidney}}, \bibinfo {author} {\bibfnamefont
  {M.}~\bibnamefont {Giustina}}, \bibinfo {author} {\bibfnamefont
  {R.}~\bibnamefont {Graff}}, \bibinfo {author} {\bibfnamefont
  {T.}~\bibnamefont {Huang}}, \bibinfo {author} {\bibfnamefont
  {E.}~\bibnamefont {Jeffrey}}, \bibinfo {author} {\bibfnamefont
  {E.}~\bibnamefont {Lucero}}, \bibinfo {author} {\bibfnamefont {J.~Y.}\
  \bibnamefont {Mutus}}, \bibinfo {author} {\bibfnamefont {O.}~\bibnamefont
  {Naaman}}, \bibinfo {author} {\bibfnamefont {C.}~\bibnamefont {Neill}},
  \bibinfo {author} {\bibfnamefont {C.}~\bibnamefont {Quintana}}, \bibinfo
  {author} {\bibfnamefont {P.}~\bibnamefont {Roushan}}, \bibinfo {author}
  {\bibfnamefont {D.}~\bibnamefont {Sank}}, \bibinfo {author} {\bibfnamefont
  {A.}~\bibnamefont {Vainsencher}}, \bibinfo {author} {\bibfnamefont
  {J.}~\bibnamefont {Wenner}}, \bibinfo {author} {\bibfnamefont {T.~C.}\
  \bibnamefont {White}}, \bibinfo {author} {\bibfnamefont {S.}~\bibnamefont
  {Boixo}}, \bibinfo {author} {\bibfnamefont {R.}~\bibnamefont {Babbush}},
  \bibinfo {author} {\bibfnamefont {V.~N.}\ \bibnamefont {Smelyanskiy}},
  \bibinfo {author} {\bibfnamefont {H.}~\bibnamefont {Neven}},\ and\ \bibinfo
  {author} {\bibfnamefont {J.~M.}\ \bibnamefont {Martinis}},\ }Fluctuations of
  energy-relaxation times in superconducting qubits,\ \href
  {https://doi.org/10.1103/PhysRevLett.121.090502} {\bibfield  {journal}
  {\bibinfo  {journal} {Phys. Rev. Lett.}\ }\textbf {\bibinfo {volume} {121}},\
  \bibinfo {pages} {090502} (\bibinfo {year} {2018})}\BibitemShut {NoStop}%
\bibitem [{\citenamefont {Burnett}\ \emph {et~al.}(2019)\citenamefont
  {Burnett}, \citenamefont {Bengtsson}, \citenamefont {Scigliuzzo},
  \citenamefont {Niepce}, \citenamefont {Kudra}, \citenamefont {Delsing},\ and\
  \citenamefont {Bylander}}]{Burnett:2019}%
  \BibitemOpen
  \bibfield  {author} {\bibinfo {author} {\bibfnamefont {J.~J.}\ \bibnamefont
  {Burnett}}, \bibinfo {author} {\bibfnamefont {A.}~\bibnamefont {Bengtsson}},
  \bibinfo {author} {\bibfnamefont {M.}~\bibnamefont {Scigliuzzo}}, \bibinfo
  {author} {\bibfnamefont {D.}~\bibnamefont {Niepce}}, \bibinfo {author}
  {\bibfnamefont {M.}~\bibnamefont {Kudra}}, \bibinfo {author} {\bibfnamefont
  {P.}~\bibnamefont {Delsing}},\ and\ \bibinfo {author} {\bibfnamefont
  {J.}~\bibnamefont {Bylander}},\ }Decoherence benchmarking of superconducting
  qubits,\ \href {https://doi.org/https://doi.org/10.1038/s41534-019-0168-5}
  {\bibfield  {journal} {\bibinfo  {journal} {npj Quantum Inf.}\ }\textbf
  {\bibinfo {volume} {5}},\ \bibinfo {pages} {54} (\bibinfo {year}
  {2019})}\BibitemShut {NoStop}%
\bibitem [{\citenamefont {Schl\"or}\ \emph {et~al.}(2019)\citenamefont
  {Schl\"or}, \citenamefont {Lisenfeld}, \citenamefont {M\"uller},
  \citenamefont {Bilmes}, \citenamefont {Schneider}, \citenamefont {Pappas},
  \citenamefont {Ustinov},\ and\ \citenamefont {Weides}}]{Schlor:2019}%
  \BibitemOpen
  \bibfield  {author} {\bibinfo {author} {\bibfnamefont {S.}~\bibnamefont
  {Schl\"or}}, \bibinfo {author} {\bibfnamefont {J.}~\bibnamefont {Lisenfeld}},
  \bibinfo {author} {\bibfnamefont {C.}~\bibnamefont {M\"uller}}, \bibinfo
  {author} {\bibfnamefont {A.}~\bibnamefont {Bilmes}}, \bibinfo {author}
  {\bibfnamefont {A.}~\bibnamefont {Schneider}}, \bibinfo {author}
  {\bibfnamefont {D.~P.}\ \bibnamefont {Pappas}}, \bibinfo {author}
  {\bibfnamefont {A.~V.}\ \bibnamefont {Ustinov}},\ and\ \bibinfo {author}
  {\bibfnamefont {M.}~\bibnamefont {Weides}},\ }Correlating decoherence in
  transmon qubits: Low frequency noise by single fluctuators,\ \href
  {https://doi.org/10.1103/PhysRevLett.123.190502} {\bibfield  {journal}
  {\bibinfo  {journal} {Phys. Rev. Lett.}\ }\textbf {\bibinfo {volume} {123}},\
  \bibinfo {pages} {190502} (\bibinfo {year} {2019})}\BibitemShut {NoStop}%
\bibitem [{\citenamefont {Lisenfeld}\ \emph {et~al.}(2019)\citenamefont
  {Lisenfeld}, \citenamefont {Bilmes}, \citenamefont {Megrant}, \citenamefont
  {Barends}, \citenamefont {Kelly}, \citenamefont {Klimov}, \citenamefont
  {Weiss}, \citenamefont {Martinis},\ and\ \citenamefont
  {Ustinov}}]{Lisenfeld:2019}%
  \BibitemOpen
  \bibfield  {author} {\bibinfo {author} {\bibfnamefont {J.}~\bibnamefont
  {Lisenfeld}}, \bibinfo {author} {\bibfnamefont {A.}~\bibnamefont {Bilmes}},
  \bibinfo {author} {\bibfnamefont {A.}~\bibnamefont {Megrant}}, \bibinfo
  {author} {\bibfnamefont {R.}~\bibnamefont {Barends}}, \bibinfo {author}
  {\bibfnamefont {J.}~\bibnamefont {Kelly}}, \bibinfo {author} {\bibfnamefont
  {P.}~\bibnamefont {Klimov}}, \bibinfo {author} {\bibfnamefont
  {G.}~\bibnamefont {Weiss}}, \bibinfo {author} {\bibfnamefont {J.~M.}\
  \bibnamefont {Martinis}},\ and\ \bibinfo {author} {\bibfnamefont {A.~V.}\
  \bibnamefont {Ustinov}},\ }Electric field spectroscopy of material defects in
  transmon qubits,\ \href
  {https://doi.org/https://doi.org/10.1038/s41534-019-0224-1} {\bibfield
  {journal} {\bibinfo  {journal} {npj Quantum Inf.}\ }\textbf {\bibinfo
  {volume} {5}},\ \bibinfo {pages} {105} (\bibinfo {year} {2019})}\BibitemShut
  {NoStop}%
\bibitem [{\citenamefont {Bilmes}\ \emph {et~al.}(2019)\citenamefont {Bilmes},
  \citenamefont {Megrant}, \citenamefont {Klimov}, \citenamefont {Weiss},
  \citenamefont {Martinis}, \citenamefont {Ustinov},\ and\ \citenamefont
  {Lisenfeld}}]{Bilmes:2019}%
  \BibitemOpen
  \bibfield  {author} {\bibinfo {author} {\bibfnamefont {A.}~\bibnamefont
  {Bilmes}}, \bibinfo {author} {\bibfnamefont {A.}~\bibnamefont {Megrant}},
  \bibinfo {author} {\bibfnamefont {P.}~\bibnamefont {Klimov}}, \bibinfo
  {author} {\bibfnamefont {G.}~\bibnamefont {Weiss}}, \bibinfo {author}
  {\bibfnamefont {J.~M.}\ \bibnamefont {Martinis}}, \bibinfo {author}
  {\bibfnamefont {A.~V.}\ \bibnamefont {Ustinov}},\ and\ \bibinfo {author}
  {\bibfnamefont {J.}~\bibnamefont {Lisenfeld}},\ }Resolving the positions of
  defects in superconducting quantum bits,\ \Eprint
  {https://arxiv.org/abs/1911.08246} {arXiv:1911.08246}  (\bibinfo {year}
  {2019})\BibitemShut {NoStop}%
\bibitem [{\citenamefont {Lehnert}\ \emph {et~al.}(1992)\citenamefont
  {Lehnert}, \citenamefont {Billon}, \citenamefont {Grassl},\ and\
  \citenamefont {Gundlach}}]{Lehnert_1992}%
  \BibitemOpen
  \bibfield  {author} {\bibinfo {author} {\bibfnamefont {T.}~\bibnamefont
  {Lehnert}}, \bibinfo {author} {\bibfnamefont {D.}~\bibnamefont {Billon}},
  \bibinfo {author} {\bibfnamefont {C.}~\bibnamefont {Grassl}},\ and\ \bibinfo
  {author} {\bibfnamefont {K.~H.}\ \bibnamefont {Gundlach}},\ }Thermal
  annealing properties of nb‐al/alox‐nb tunnel junctions,\ \href
  {https://doi.org/https://doi.org/10.1063/1.351479} {\bibfield  {journal}
  {\bibinfo  {journal} {Journal of Applied Physics}\ }\textbf {\bibinfo
  {volume} {72}},\ \bibinfo {pages} {3165} (\bibinfo {year}
  {1992})}\BibitemShut {NoStop}%
\bibitem [{\citenamefont {Oliva}\ and\ \citenamefont
  {Monaco}(1994)}]{Oliva_1994}%
  \BibitemOpen
  \bibfield  {author} {\bibinfo {author} {\bibfnamefont {A.}~\bibnamefont
  {Oliva}}\ and\ \bibinfo {author} {\bibfnamefont {R.}~\bibnamefont {Monaco}},\
  }Annealing properties of high quality $\rm{Nb}/{Al}{-}{AlO}_{x}/{Nb}$ tunnel
  junctions,\ \href {https://doi.org/https://doi.org/10.1109/77.273061}
  {\bibfield  {journal} {\bibinfo  {journal} {IEEE Transactions on Applied
  Superconductivity}\ }\textbf {\bibinfo {volume} {4}},\ \bibinfo {pages} {25}
  (\bibinfo {year} {1994})}\BibitemShut {NoStop}%
\bibitem [{\citenamefont {Potts}\ \emph {et~al.}(2001)\citenamefont {Potts},
  \citenamefont {Parker}, \citenamefont {Baumberg},\ and\ \citenamefont
  {Groot}}]{Potts_2001}%
  \BibitemOpen
  \bibfield  {author} {\bibinfo {author} {\bibfnamefont {A.}~\bibnamefont
  {Potts}}, \bibinfo {author} {\bibfnamefont {G.}~\bibnamefont {Parker}},
  \bibinfo {author} {\bibfnamefont {J.}~\bibnamefont {Baumberg}},\ and\
  \bibinfo {author} {\bibfnamefont {P.}~\bibnamefont {Groot}},\ }Cmos
  compatible fabrication methods for submicron josephson junction qubits,\
  \href {https://doi.org/10.1049/ip-smt:20010395} {\bibfield  {journal}
  {\bibinfo  {journal} {Science, Measurement and Technology, IEE Proceedings
  -}\ }\textbf {\bibinfo {volume} {148}},\ \bibinfo {pages} {225 } (\bibinfo
  {year} {2001})}\BibitemShut {NoStop}%
\bibitem [{\citenamefont {Koppinen}\ \emph {et~al.}(2007)\citenamefont
  {Koppinen}, \citenamefont {Vaisto},\ and\ \citenamefont
  {Maasilta}}]{Koppinen_2007}%
  \BibitemOpen
  \bibfield  {author} {\bibinfo {author} {\bibfnamefont {P.}~\bibnamefont
  {Koppinen}}, \bibinfo {author} {\bibfnamefont {L.~M.}\ \bibnamefont
  {Vaisto}},\ and\ \bibinfo {author} {\bibfnamefont {I.}~\bibnamefont
  {Maasilta}},\ }Complete stabilization and improvement of the characteristics
  of tunnel junctions by thermal annealing,\ \href
  {https://doi.org/https://doi.org/10.1063/1.2437662} {\bibfield  {journal}
  {\bibinfo  {journal} {Applied Physics Letters}\ }\textbf {\bibinfo {volume}
  {90}},\ \bibinfo {pages} {053503} (\bibinfo {year} {2007})}\BibitemShut
  {NoStop}%
\bibitem [{\citenamefont {Granata}\ \emph {et~al.}(2008)\citenamefont
  {Granata}, \citenamefont {Vettoliere}, \citenamefont {Petti}, \citenamefont
  {Rippa}, \citenamefont {Ruggiero}, \citenamefont {Mormile},\ and\
  \citenamefont {Russo}}]{Granata_2008}%
  \BibitemOpen
  \bibfield  {author} {\bibinfo {author} {\bibfnamefont {C.}~\bibnamefont
  {Granata}}, \bibinfo {author} {\bibfnamefont {A.}~\bibnamefont {Vettoliere}},
  \bibinfo {author} {\bibfnamefont {L.}~\bibnamefont {Petti}}, \bibinfo
  {author} {\bibfnamefont {M.}~\bibnamefont {Rippa}}, \bibinfo {author}
  {\bibfnamefont {B.}~\bibnamefont {Ruggiero}}, \bibinfo {author}
  {\bibfnamefont {P.}~\bibnamefont {Mormile}},\ and\ \bibinfo {author}
  {\bibfnamefont {M.}~\bibnamefont {Russo}},\ }Trimming of critical current in
  niobium josephson devices by laser annealing,\ \href
  {https://doi.org/10.1088/1742-6596/97/1/012110} {\bibfield  {journal}
  {\bibinfo  {journal} {Journal of Physics: Conference Series}\ }\textbf
  {\bibinfo {volume} {97}},\ \bibinfo {pages} {012110} (\bibinfo {year}
  {2008})}\BibitemShut {NoStop}%
\bibitem [{\citenamefont {Bumble}\ \emph {et~al.}(2009)\citenamefont {Bumble},
  \citenamefont {Fung}, \citenamefont {Kaul}, \citenamefont {Kleinsasser},
  \citenamefont {Kerber}, \citenamefont {Bunyk},\ and\ \citenamefont
  {Ladizinsky}}]{Bumble_2009}%
  \BibitemOpen
  \bibfield  {author} {\bibinfo {author} {\bibfnamefont {B.}~\bibnamefont
  {Bumble}}, \bibinfo {author} {\bibfnamefont {A.}~\bibnamefont {Fung}},
  \bibinfo {author} {\bibfnamefont {A.}~\bibnamefont {Kaul}}, \bibinfo {author}
  {\bibfnamefont {A.~W.}\ \bibnamefont {Kleinsasser}}, \bibinfo {author}
  {\bibfnamefont {G.~L.}\ \bibnamefont {Kerber}}, \bibinfo {author}
  {\bibfnamefont {P.}~\bibnamefont {Bunyk}},\ and\ \bibinfo {author}
  {\bibfnamefont {E.}~\bibnamefont {Ladizinsky}},\ }Submicrometer
  $\rm{Nb}/{Al}{-}{AlO}_{x}/{Nb}$ integrated circuit fabrication process for
  quantum computing applications,\ \href
  {https://doi.org/10.1109/TASC.2009.2018249} {\bibfield  {journal} {\bibinfo
  {journal} {IEEE Transactions on Applied Superconductivity}\ }\textbf
  {\bibinfo {volume} {19}},\ \bibinfo {pages} {226} (\bibinfo {year}
  {2009})}\BibitemShut {NoStop}%
\bibitem [{\citenamefont {Pop}\ \emph {et~al.}(2012)\citenamefont {Pop},
  \citenamefont {Fournier}, \citenamefont {Crozes}, \citenamefont {Lecocq},
  \citenamefont {Matei}, \citenamefont {Pannetier}, \citenamefont {Buisson},\
  and\ \citenamefont {Guichard}}]{Pop_2012}%
  \BibitemOpen
  \bibfield  {author} {\bibinfo {author} {\bibfnamefont {I.}~\bibnamefont
  {Pop}}, \bibinfo {author} {\bibfnamefont {T.}~\bibnamefont {Fournier}},
  \bibinfo {author} {\bibfnamefont {T.}~\bibnamefont {Crozes}}, \bibinfo
  {author} {\bibfnamefont {F.}~\bibnamefont {Lecocq}}, \bibinfo {author}
  {\bibfnamefont {I.}~\bibnamefont {Matei}}, \bibinfo {author} {\bibfnamefont
  {B.}~\bibnamefont {Pannetier}}, \bibinfo {author} {\bibfnamefont
  {O.}~\bibnamefont {Buisson}},\ and\ \bibinfo {author} {\bibfnamefont
  {W.}~\bibnamefont {Guichard}},\ }Fabrication of stable and reproducible
  submicron tunnel junctions,\ \href
  {https://doi.org/https://doi.org/10.1116/1.3673790} {\bibfield  {journal}
  {\bibinfo  {journal} {Journal of Vacuum Science \& Technology B}\ }\textbf
  {\bibinfo {volume} {30}},\ \bibinfo {pages} {010607} (\bibinfo {year}
  {2012})}\BibitemShut {NoStop}%
\bibitem [{\citenamefont {Costache}\ \emph {et~al.}(2012)\citenamefont
  {Costache}, \citenamefont {Bridoux}, \citenamefont {Neumann},\ and\
  \citenamefont {Valenzuela}}]{Costache_2012}%
  \BibitemOpen
  \bibfield  {author} {\bibinfo {author} {\bibfnamefont {M.~V.}\ \bibnamefont
  {Costache}}, \bibinfo {author} {\bibfnamefont {G.}~\bibnamefont {Bridoux}},
  \bibinfo {author} {\bibfnamefont {I.}~\bibnamefont {Neumann}},\ and\ \bibinfo
  {author} {\bibfnamefont {S.~O.}\ \bibnamefont {Valenzuela}},\ }Lateral
  metallic devices made by a multiangle shadow evaporation technique,\ \href
  {https://doi.org/10.1116/1.4722982} {\bibfield  {journal} {\bibinfo
  {journal} {Journal of Vacuum Science \& Technology B}\ }\textbf {\bibinfo
  {volume} {30}},\ \bibinfo {pages} {04E105} (\bibinfo {year}
  {2012})}\BibitemShut {NoStop}%
\bibitem [{\citenamefont {Wu}\ \emph {et~al.}(2017)\citenamefont {Wu},
  \citenamefont {Long}, \citenamefont {Ku}, \citenamefont {Lake}, \citenamefont
  {Bal},\ and\ \citenamefont {Pappas}}]{Wu_2017}%
  \BibitemOpen
  \bibfield  {author} {\bibinfo {author} {\bibfnamefont {X.}~\bibnamefont
  {Wu}}, \bibinfo {author} {\bibfnamefont {J.~L.}\ \bibnamefont {Long}},
  \bibinfo {author} {\bibfnamefont {H.~S.}\ \bibnamefont {Ku}}, \bibinfo
  {author} {\bibfnamefont {R.~E.}\ \bibnamefont {Lake}}, \bibinfo {author}
  {\bibfnamefont {M.}~\bibnamefont {Bal}},\ and\ \bibinfo {author}
  {\bibfnamefont {D.~P.}\ \bibnamefont {Pappas}},\ }Overlap junctions for high
  coherence superconducting qubits,\ \href {https://doi.org/10.1063/1.4993937}
  {\bibfield  {journal} {\bibinfo  {journal} {Applied Physics Letters}\
  }\textbf {\bibinfo {volume} {111}},\ \bibinfo {pages} {032602} (\bibinfo
  {year} {2017})}\BibitemShut {NoStop}%
\bibitem [{\citenamefont {Brink}\ \emph {et~al.}(2018)\citenamefont {Brink},
  \citenamefont {Chow}, \citenamefont {Hertzberg}, \citenamefont {Magesan},\
  and\ \citenamefont {Rosenblatt}}]{Brink:2018}%
  \BibitemOpen
  \bibfield  {author} {\bibinfo {author} {\bibfnamefont {M.}~\bibnamefont
  {Brink}}, \bibinfo {author} {\bibfnamefont {J.}~\bibnamefont {Chow}},
  \bibinfo {author} {\bibfnamefont {J.}~\bibnamefont {Hertzberg}}, \bibinfo
  {author} {\bibfnamefont {E.}~\bibnamefont {Magesan}},\ and\ \bibinfo {author}
  {\bibfnamefont {S.}~\bibnamefont {Rosenblatt}},\ }Device challenges for near
  term superconducting quantum processors: frequency collisions,\ \href
  {https://doi.org/10.1109/IEDM.2018.8614500} {\bibfield  {journal} {\bibinfo
  {journal} {2018 IEEE International Electron Devices Meeting (IEDM)}\ ,\
  \bibinfo {pages} {6.1.1}} (\bibinfo {year} {2018})}\BibitemShut {NoStop}%
\bibitem [{\citenamefont {Kreikebaum}\ \emph {et~al.}(2020)\citenamefont
  {Kreikebaum}, \citenamefont {O'Brien}, \citenamefont {Morvan},\ and\
  \citenamefont {Siddiqi}}]{Kreikebaum_2020}%
  \BibitemOpen
  \bibfield  {author} {\bibinfo {author} {\bibfnamefont {J.~M.}\ \bibnamefont
  {Kreikebaum}}, \bibinfo {author} {\bibfnamefont {K.~P.}\ \bibnamefont
  {O'Brien}}, \bibinfo {author} {\bibfnamefont {A.}~\bibnamefont {Morvan}},\
  and\ \bibinfo {author} {\bibfnamefont {I.}~\bibnamefont {Siddiqi}},\
  }Improving wafer-scale josephson junction resistance variation in
  superconducting quantum coherent circuits,\ \href
  {https://doi.org/10.1088/1361-6668/ab8617} {\bibfield  {journal} {\bibinfo
  {journal} {Superconductor Science and Technology}\ }\textbf {\bibinfo
  {volume} {33}},\ \bibinfo {pages} {06LT02} (\bibinfo {year}
  {2020})}\BibitemShut {NoStop}%
\bibitem [{\citenamefont {Dickel}\ \emph {et~al.}(2018)\citenamefont {Dickel},
  \citenamefont {Wesdorp}, \citenamefont {Langford}, \citenamefont {Peiter},
  \citenamefont {Sagastizabal}, \citenamefont {Bruno}, \citenamefont {Criger},
  \citenamefont {Motzoi},\ and\ \citenamefont {DiCarlo}}]{Dickel:2018}%
  \BibitemOpen
  \bibfield  {author} {\bibinfo {author} {\bibfnamefont {C.}~\bibnamefont
  {Dickel}}, \bibinfo {author} {\bibfnamefont {J.~J.}\ \bibnamefont {Wesdorp}},
  \bibinfo {author} {\bibfnamefont {N.~K.}\ \bibnamefont {Langford}}, \bibinfo
  {author} {\bibfnamefont {S.}~\bibnamefont {Peiter}}, \bibinfo {author}
  {\bibfnamefont {R.}~\bibnamefont {Sagastizabal}}, \bibinfo {author}
  {\bibfnamefont {A.}~\bibnamefont {Bruno}}, \bibinfo {author} {\bibfnamefont
  {B.}~\bibnamefont {Criger}}, \bibinfo {author} {\bibfnamefont
  {F.}~\bibnamefont {Motzoi}},\ and\ \bibinfo {author} {\bibfnamefont
  {L.}~\bibnamefont {DiCarlo}},\ }Chip-to-chip entanglement of transmon qubits
  using engineered measurement fields,\ \href
  {http://dx.doi.org/10.1103/PhysRevB.97.064508} {\bibfield  {journal}
  {\bibinfo  {journal} {Physical Review B}\ }\textbf {\bibinfo {volume} {97}}
  (\bibinfo {year} {2018})}\BibitemShut {NoStop}%
\bibitem [{\citenamefont {Chou}\ \emph {et~al.}(2018)\citenamefont {Chou},
  \citenamefont {Blumoff}, \citenamefont {Wang}, \citenamefont {Reinhold},
  \citenamefont {Axline}, \citenamefont {Gao}, \citenamefont {Frunzio},
  \citenamefont {Devoret}, \citenamefont {Jiang},\ and\ \citenamefont
  {Schoelkopf}}]{Chou:2018}%
  \BibitemOpen
  \bibfield  {author} {\bibinfo {author} {\bibfnamefont {K.~S.}\ \bibnamefont
  {Chou}}, \bibinfo {author} {\bibfnamefont {J.~Z.}\ \bibnamefont {Blumoff}},
  \bibinfo {author} {\bibfnamefont {C.~S.}\ \bibnamefont {Wang}}, \bibinfo
  {author} {\bibfnamefont {P.~C.}\ \bibnamefont {Reinhold}}, \bibinfo {author}
  {\bibfnamefont {C.~J.}\ \bibnamefont {Axline}}, \bibinfo {author}
  {\bibfnamefont {Y.~Y.}\ \bibnamefont {Gao}}, \bibinfo {author} {\bibfnamefont
  {L.}~\bibnamefont {Frunzio}}, \bibinfo {author} {\bibfnamefont {M.~H.}\
  \bibnamefont {Devoret}}, \bibinfo {author} {\bibfnamefont {L.}~\bibnamefont
  {Jiang}},\ and\ \bibinfo {author} {\bibfnamefont {R.~J.}\ \bibnamefont
  {Schoelkopf}},\ }Deterministic teleportation of a quantum gate between two
  logical qubits,\ \href {https://doi.org/10.1038/s41586-018-0470-y} {\bibfield
   {journal} {\bibinfo  {journal} {Nature}\ }\textbf {\bibinfo {volume}
  {561}},\ \bibinfo {pages} {368–373} (\bibinfo {year} {2018})}\BibitemShut
  {NoStop}%
\bibitem [{\citenamefont {Zhong}\ \emph {et~al.}(2021)\citenamefont {Zhong},
  \citenamefont {Chang}, \citenamefont {Bienfait}, \citenamefont {Dumur},
  \citenamefont {Chou}, \citenamefont {Conner}, \citenamefont {Grebel},
  \citenamefont {Povey}, \citenamefont {Yan}, \citenamefont {Schuster},\ and\
  \citenamefont {Cleland}}]{Zhong:2020}%
  \BibitemOpen
  \bibfield  {author} {\bibinfo {author} {\bibfnamefont {Y.}~\bibnamefont
  {Zhong}}, \bibinfo {author} {\bibfnamefont {H.-S.}\ \bibnamefont {Chang}},
  \bibinfo {author} {\bibfnamefont {A.}~\bibnamefont {Bienfait}}, \bibinfo
  {author} {\bibfnamefont {{\'E}.}~\bibnamefont {Dumur}}, \bibinfo {author}
  {\bibfnamefont {M.-H.}\ \bibnamefont {Chou}}, \bibinfo {author}
  {\bibfnamefont {C.~R.}\ \bibnamefont {Conner}}, \bibinfo {author}
  {\bibfnamefont {J.}~\bibnamefont {Grebel}}, \bibinfo {author} {\bibfnamefont
  {R.~G.}\ \bibnamefont {Povey}}, \bibinfo {author} {\bibfnamefont
  {H.}~\bibnamefont {Yan}}, \bibinfo {author} {\bibfnamefont {D.~I.}\
  \bibnamefont {Schuster}},\ and\ \bibinfo {author} {\bibfnamefont {A.~N.}\
  \bibnamefont {Cleland}},\ }Deterministic multi-qubit entanglement in a
  quantum network,\ \href {https://doi.org/10.1038/s41586-021-03288-7}
  {\bibfield  {journal} {\bibinfo  {journal} {Nature}\ }\textbf {\bibinfo
  {volume} {590}},\ \bibinfo {pages} {571} (\bibinfo {year}
  {2021})}\BibitemShut {NoStop}%
\bibitem [{\citenamefont {Gold}\ \emph {et~al.}(2021)\citenamefont {Gold},
  \citenamefont {Paquette}, \citenamefont {Stockklauser}, \citenamefont
  {Reagor}, \citenamefont {Alam}, \citenamefont {Bestwick}, \citenamefont
  {Didier}, \citenamefont {Nersisyan}, \citenamefont {Oruc}, \citenamefont
  {Razavi}, \citenamefont {Scharmann}, \citenamefont {Sete}, \citenamefont
  {Sur}, \citenamefont {Venturelli}, \citenamefont {Winkleblack}, \citenamefont
  {Wudarski}, \citenamefont {Harburn},\ and\ \citenamefont
  {Rigetti}}]{Gold_2021}%
  \BibitemOpen
  \bibfield  {author} {\bibinfo {author} {\bibfnamefont {A.}~\bibnamefont
  {Gold}}, \bibinfo {author} {\bibfnamefont {J.}~\bibnamefont {Paquette}},
  \bibinfo {author} {\bibfnamefont {A.}~\bibnamefont {Stockklauser}}, \bibinfo
  {author} {\bibfnamefont {M.~J.}\ \bibnamefont {Reagor}}, \bibinfo {author}
  {\bibfnamefont {M.~S.}\ \bibnamefont {Alam}}, \bibinfo {author}
  {\bibfnamefont {A.}~\bibnamefont {Bestwick}}, \bibinfo {author}
  {\bibfnamefont {N.}~\bibnamefont {Didier}}, \bibinfo {author} {\bibfnamefont
  {A.}~\bibnamefont {Nersisyan}}, \bibinfo {author} {\bibfnamefont
  {F.}~\bibnamefont {Oruc}}, \bibinfo {author} {\bibfnamefont {A.}~\bibnamefont
  {Razavi}}, \bibinfo {author} {\bibfnamefont {B.}~\bibnamefont {Scharmann}},
  \bibinfo {author} {\bibfnamefont {E.~A.}\ \bibnamefont {Sete}}, \bibinfo
  {author} {\bibfnamefont {B.}~\bibnamefont {Sur}}, \bibinfo {author}
  {\bibfnamefont {D.}~\bibnamefont {Venturelli}}, \bibinfo {author}
  {\bibfnamefont {C.~J.}\ \bibnamefont {Winkleblack}}, \bibinfo {author}
  {\bibfnamefont {F.}~\bibnamefont {Wudarski}}, \bibinfo {author}
  {\bibfnamefont {M.}~\bibnamefont {Harburn}},\ and\ \bibinfo {author}
  {\bibfnamefont {C.}~\bibnamefont {Rigetti}},\ }Entanglement across separate
  silicon dies in a modular superconducting qubit device,\ \Eprint
  {https://arxiv.org/abs/2102.13293} {arXiv:2102.13293 [quant-ph]}  (\bibinfo
  {year} {2021})\BibitemShut {NoStop}%
\bibitem [{\citenamefont {Koch}\ \emph {et~al.}(1983)\citenamefont {Koch},
  \citenamefont {Clarke}, \citenamefont {Goubau}, \citenamefont {Martinis},
  \citenamefont {Pegrum},\ and\ \citenamefont {van Harlingen}}]{Koch:1983}%
  \BibitemOpen
  \bibfield  {author} {\bibinfo {author} {\bibfnamefont {R.~H.}\ \bibnamefont
  {Koch}}, \bibinfo {author} {\bibfnamefont {J.}~\bibnamefont {Clarke}},
  \bibinfo {author} {\bibfnamefont {W.~M.}\ \bibnamefont {Goubau}}, \bibinfo
  {author} {\bibfnamefont {J.~M.}\ \bibnamefont {Martinis}}, \bibinfo {author}
  {\bibfnamefont {C.~M.}\ \bibnamefont {Pegrum}},\ and\ \bibinfo {author}
  {\bibfnamefont {D.~J.}\ \bibnamefont {van Harlingen}},\ }Flicker ($1/f$)
  noise in tunnel junction dc {SQUID}s,\ \href
  {https://doi.org/https://doi.org/10.1007/BF00683423} {\bibfield  {journal}
  {\bibinfo  {journal} {J. Low Temp. Phys.}\ }\textbf {\bibinfo {volume}
  {51}},\ \bibinfo {pages} {207} (\bibinfo {year} {1983})}\BibitemShut
  {NoStop}%
\bibitem [{\citenamefont {Wellstood}\ \emph {et~al.}(1987)\citenamefont
  {Wellstood}, \citenamefont {Urbina},\ and\ \citenamefont
  {Clarke}}]{Wellstood:1987}%
  \BibitemOpen
  \bibfield  {author} {\bibinfo {author} {\bibfnamefont {F.~C.}\ \bibnamefont
  {Wellstood}}, \bibinfo {author} {\bibfnamefont {C.}~\bibnamefont {Urbina}},\
  and\ \bibinfo {author} {\bibfnamefont {J.}~\bibnamefont {Clarke}},\
  }Low-frequency noise in dc superconducting quantum interference devices below
  1 {K},\ \href {https://doi.org/https://doi.org/10.1063/1.98041} {\bibfield
  {journal} {\bibinfo  {journal} {Appl. Phys. Lett.}\ }\textbf {\bibinfo
  {volume} {50}},\ \bibinfo {pages} {772} (\bibinfo {year} {1987})}\BibitemShut
  {NoStop}%
\bibitem [{\citenamefont {Vion}\ \emph {et~al.}(2002)\citenamefont {Vion},
  \citenamefont {Aassime}, \citenamefont {Cottet}, \citenamefont {Joyez},
  \citenamefont {Pothier}, \citenamefont {Urbina}, \citenamefont {Esteve},\
  and\ \citenamefont {Devoret}}]{Vion:2002}%
  \BibitemOpen
  \bibfield  {author} {\bibinfo {author} {\bibfnamefont {D.}~\bibnamefont
  {Vion}}, \bibinfo {author} {\bibfnamefont {A.}~\bibnamefont {Aassime}},
  \bibinfo {author} {\bibfnamefont {A.}~\bibnamefont {Cottet}}, \bibinfo
  {author} {\bibfnamefont {P.}~\bibnamefont {Joyez}}, \bibinfo {author}
  {\bibfnamefont {H.}~\bibnamefont {Pothier}}, \bibinfo {author} {\bibfnamefont
  {C.}~\bibnamefont {Urbina}}, \bibinfo {author} {\bibfnamefont
  {D.}~\bibnamefont {Esteve}},\ and\ \bibinfo {author} {\bibfnamefont {M.~H.}\
  \bibnamefont {Devoret}},\ }Manipulating the quantum state of an electrical
  circuit,\ \href {https://doi.org/10.1126/science.1069372} {\bibfield
  {journal} {\bibinfo  {journal} {Science}\ }\textbf {\bibinfo {volume}
  {296}},\ \bibinfo {pages} {886} (\bibinfo {year} {2002})}\BibitemShut
  {NoStop}%
\bibitem [{\citenamefont {Martinis}\ \emph {et~al.}(2003)\citenamefont
  {Martinis}, \citenamefont {Nam}, \citenamefont {Aumentado}, \citenamefont
  {Lang},\ and\ \citenamefont {Urbina}}]{Martinis:2003}%
  \BibitemOpen
  \bibfield  {author} {\bibinfo {author} {\bibfnamefont {J.~M.}\ \bibnamefont
  {Martinis}}, \bibinfo {author} {\bibfnamefont {S.}~\bibnamefont {Nam}},
  \bibinfo {author} {\bibfnamefont {J.}~\bibnamefont {Aumentado}}, \bibinfo
  {author} {\bibfnamefont {K.~M.}\ \bibnamefont {Lang}},\ and\ \bibinfo
  {author} {\bibfnamefont {C.}~\bibnamefont {Urbina}},\ }Decoherence of a
  superconducting qubit due to bias noise,\ \href
  {https://doi.org/10.1103/PhysRevB.67.094510} {\bibfield  {journal} {\bibinfo
  {journal} {Phys. Rev. B}\ }\textbf {\bibinfo {volume} {67}},\ \bibinfo
  {pages} {094510} (\bibinfo {year} {2003})}\BibitemShut {NoStop}%
\bibitem [{\citenamefont {Ithier}\ \emph {et~al.}(2005)\citenamefont {Ithier},
  \citenamefont {Collin}, \citenamefont {Joyez}, \citenamefont {Meeson},
  \citenamefont {Vion}, \citenamefont {Esteve}, \citenamefont {Chiarello},
  \citenamefont {Shnirman}, \citenamefont {Makhlin}, \citenamefont {Schriefl},\
  and\ \citenamefont {Sch\"on}}]{Ithier:2005}%
  \BibitemOpen
  \bibfield  {author} {\bibinfo {author} {\bibfnamefont {G.}~\bibnamefont
  {Ithier}}, \bibinfo {author} {\bibfnamefont {E.}~\bibnamefont {Collin}},
  \bibinfo {author} {\bibfnamefont {P.}~\bibnamefont {Joyez}}, \bibinfo
  {author} {\bibfnamefont {P.~J.}\ \bibnamefont {Meeson}}, \bibinfo {author}
  {\bibfnamefont {D.}~\bibnamefont {Vion}}, \bibinfo {author} {\bibfnamefont
  {D.}~\bibnamefont {Esteve}}, \bibinfo {author} {\bibfnamefont
  {F.}~\bibnamefont {Chiarello}}, \bibinfo {author} {\bibfnamefont
  {A.}~\bibnamefont {Shnirman}}, \bibinfo {author} {\bibfnamefont
  {Y.}~\bibnamefont {Makhlin}}, \bibinfo {author} {\bibfnamefont
  {J.}~\bibnamefont {Schriefl}},\ and\ \bibinfo {author} {\bibfnamefont
  {G.}~\bibnamefont {Sch\"on}},\ }Decoherence in a superconducting quantum bit
  circuit,\ \href {https://doi.org/https://doi.org/10.1103/PhysRevB.72.134519}
  {\bibfield  {journal} {\bibinfo  {journal} {Phys. Rev. B}\ }\textbf {\bibinfo
  {volume} {72}},\ \bibinfo {pages} {134519} (\bibinfo {year}
  {2005})}\BibitemShut {NoStop}%
\bibitem [{\citenamefont {Yoshihara}\ \emph {et~al.}(2006)\citenamefont
  {Yoshihara}, \citenamefont {Harrabi}, \citenamefont {Niskanen}, \citenamefont
  {Nakamura},\ and\ \citenamefont {Tsai}}]{Yoshihara:2006}%
  \BibitemOpen
  \bibfield  {author} {\bibinfo {author} {\bibfnamefont {F.}~\bibnamefont
  {Yoshihara}}, \bibinfo {author} {\bibfnamefont {K.}~\bibnamefont {Harrabi}},
  \bibinfo {author} {\bibfnamefont {A.~O.}\ \bibnamefont {Niskanen}}, \bibinfo
  {author} {\bibfnamefont {Y.}~\bibnamefont {Nakamura}},\ and\ \bibinfo
  {author} {\bibfnamefont {J.~S.}\ \bibnamefont {Tsai}},\ }Decoherence of flux
  qubits due to $1/f$ flux noise,\ \href
  {https://doi.org/10.1103/PhysRevLett.97.167001} {\bibfield  {journal}
  {\bibinfo  {journal} {Phys. Rev. Lett.}\ }\textbf {\bibinfo {volume} {97}},\
  \bibinfo {pages} {167001} (\bibinfo {year} {2006})}\BibitemShut {NoStop}%
\bibitem [{\citenamefont {Bialczak}\ \emph {et~al.}(2007)\citenamefont
  {Bialczak}, \citenamefont {McDermott}, \citenamefont {Ansmann}, \citenamefont
  {Hofheinz}, \citenamefont {Katz}, \citenamefont {Lucero}, \citenamefont
  {Neeley}, \citenamefont {O'Connell}, \citenamefont {Wang}, \citenamefont
  {Cleland},\ and\ \citenamefont {Martinis}}]{Bialczak:2007}%
  \BibitemOpen
  \bibfield  {author} {\bibinfo {author} {\bibfnamefont {R.~C.}\ \bibnamefont
  {Bialczak}}, \bibinfo {author} {\bibfnamefont {R.}~\bibnamefont {McDermott}},
  \bibinfo {author} {\bibfnamefont {M.}~\bibnamefont {Ansmann}}, \bibinfo
  {author} {\bibfnamefont {M.}~\bibnamefont {Hofheinz}}, \bibinfo {author}
  {\bibfnamefont {N.}~\bibnamefont {Katz}}, \bibinfo {author} {\bibfnamefont
  {E.}~\bibnamefont {Lucero}}, \bibinfo {author} {\bibfnamefont
  {M.}~\bibnamefont {Neeley}}, \bibinfo {author} {\bibfnamefont {A.~D.}\
  \bibnamefont {O'Connell}}, \bibinfo {author} {\bibfnamefont {H.}~\bibnamefont
  {Wang}}, \bibinfo {author} {\bibfnamefont {A.~N.}\ \bibnamefont {Cleland}},\
  and\ \bibinfo {author} {\bibfnamefont {J.~M.}\ \bibnamefont {Martinis}},\
  }$1/f$ flux noise in josephson phase qubits,\ \href
  {https://doi.org/10.1103/PhysRevLett.99.187006} {\bibfield  {journal}
  {\bibinfo  {journal} {Phys. Rev. Lett.}\ }\textbf {\bibinfo {volume} {99}},\
  \bibinfo {pages} {187006} (\bibinfo {year} {2007})}\BibitemShut {NoStop}%
\bibitem [{\citenamefont {Koch}\ \emph
  {et~al.}(2007{\natexlab{a}})\citenamefont {Koch}, \citenamefont
  {DiVincenzo},\ and\ \citenamefont {Clarke}}]{KochR:2007}%
  \BibitemOpen
  \bibfield  {author} {\bibinfo {author} {\bibfnamefont {R.~H.}\ \bibnamefont
  {Koch}}, \bibinfo {author} {\bibfnamefont {D.~P.}\ \bibnamefont
  {DiVincenzo}},\ and\ \bibinfo {author} {\bibfnamefont {J.}~\bibnamefont
  {Clarke}},\ }Model for $1/f$ flux noise in {SQUID}s and qubits,\ \href
  {https://doi.org/10.1103/PhysRevLett.98.267003} {\bibfield  {journal}
  {\bibinfo  {journal} {Phys. Rev. Lett.}\ }\textbf {\bibinfo {volume} {98}},\
  \bibinfo {pages} {267003} (\bibinfo {year} {2007}{\natexlab{a}})}\BibitemShut
  {NoStop}%
\bibitem [{\citenamefont {Faoro}\ and\ \citenamefont
  {Ioffe}(2008)}]{Faoro:2008}%
  \BibitemOpen
  \bibfield  {author} {\bibinfo {author} {\bibfnamefont {L.}~\bibnamefont
  {Faoro}}\ and\ \bibinfo {author} {\bibfnamefont {L.~B.}\ \bibnamefont
  {Ioffe}},\ }Microscopic origin of low-frequency flux noise in josephson
  circuits,\ \href {https://doi.org/10.1103/PhysRevLett.100.227005} {\bibfield
  {journal} {\bibinfo  {journal} {Phys. Rev. Lett.}\ }\textbf {\bibinfo
  {volume} {100}},\ \bibinfo {pages} {227005} (\bibinfo {year}
  {2008})}\BibitemShut {NoStop}%
\bibitem [{\citenamefont {Manucharyan}\ \emph {et~al.}(2009)\citenamefont
  {Manucharyan}, \citenamefont {Koch}, \citenamefont {Glazman},\ and\
  \citenamefont {Devoret}}]{Manucharyan:2009}%
  \BibitemOpen
  \bibfield  {author} {\bibinfo {author} {\bibfnamefont {V.~E.}\ \bibnamefont
  {Manucharyan}}, \bibinfo {author} {\bibfnamefont {J.}~\bibnamefont {Koch}},
  \bibinfo {author} {\bibfnamefont {L.~I.}\ \bibnamefont {Glazman}},\ and\
  \bibinfo {author} {\bibfnamefont {M.~H.}\ \bibnamefont {Devoret}},\
  }Fluxonium: Single cooper-pair circuit free of charge offsets,\ \href
  {https://doi.org/10.1126/science.1175552} {\bibfield  {journal} {\bibinfo
  {journal} {Science}\ }\textbf {\bibinfo {volume} {326}},\ \bibinfo {pages}
  {113} (\bibinfo {year} {2009})}\BibitemShut {NoStop}%
\bibitem [{\citenamefont {Barends}\ \emph {et~al.}(2013)\citenamefont
  {Barends}, \citenamefont {Kelly}, \citenamefont {Megrant}, \citenamefont
  {Sank}, \citenamefont {Jeffrey}, \citenamefont {Chen}, \citenamefont {Yin},
  \citenamefont {Chiaro}, \citenamefont {Mutus}, \citenamefont {Neill},
  \citenamefont {O'Malley}, \citenamefont {Roushan}, \citenamefont {Wenner},
  \citenamefont {White}, \citenamefont {Cleland},\ and\ \citenamefont
  {Martinis}}]{Barends:2013}%
  \BibitemOpen
  \bibfield  {author} {\bibinfo {author} {\bibfnamefont {R.}~\bibnamefont
  {Barends}}, \bibinfo {author} {\bibfnamefont {J.}~\bibnamefont {Kelly}},
  \bibinfo {author} {\bibfnamefont {A.}~\bibnamefont {Megrant}}, \bibinfo
  {author} {\bibfnamefont {D.}~\bibnamefont {Sank}}, \bibinfo {author}
  {\bibfnamefont {E.}~\bibnamefont {Jeffrey}}, \bibinfo {author} {\bibfnamefont
  {Y.}~\bibnamefont {Chen}}, \bibinfo {author} {\bibfnamefont {Y.}~\bibnamefont
  {Yin}}, \bibinfo {author} {\bibfnamefont {B.}~\bibnamefont {Chiaro}},
  \bibinfo {author} {\bibfnamefont {J.}~\bibnamefont {Mutus}}, \bibinfo
  {author} {\bibfnamefont {C.}~\bibnamefont {Neill}}, \bibinfo {author}
  {\bibfnamefont {P.}~\bibnamefont {O'Malley}}, \bibinfo {author}
  {\bibfnamefont {P.}~\bibnamefont {Roushan}}, \bibinfo {author} {\bibfnamefont
  {J.}~\bibnamefont {Wenner}}, \bibinfo {author} {\bibfnamefont {T.~C.}\
  \bibnamefont {White}}, \bibinfo {author} {\bibfnamefont {A.~N.}\ \bibnamefont
  {Cleland}},\ and\ \bibinfo {author} {\bibfnamefont {J.~M.}\ \bibnamefont
  {Martinis}},\ }Coherent josephson qubit suitable for scalable quantum
  integrated circuits,\ \href {https://doi.org/10.1103/PhysRevLett.111.080502}
  {\bibfield  {journal} {\bibinfo  {journal} {Phys. Rev. Lett.}\ }\textbf
  {\bibinfo {volume} {111}},\ \bibinfo {pages} {080502} (\bibinfo {year}
  {2013})}\BibitemShut {NoStop}%
\bibitem [{\citenamefont {O'Malley}\ \emph {et~al.}(2015)\citenamefont
  {O'Malley}, \citenamefont {Kelly}, \citenamefont {Barends}, \citenamefont
  {Campbell}, \citenamefont {Chen}, \citenamefont {Chen}, \citenamefont
  {Chiaro}, \citenamefont {Dunsworth}, \citenamefont {Fowler}, \citenamefont
  {Hoi}, \citenamefont {Jeffrey}, \citenamefont {Megrant}, \citenamefont
  {Mutus}, \citenamefont {Neill}, \citenamefont {Quintana}, \citenamefont
  {Roushan}, \citenamefont {Sank}, \citenamefont {Vainsencher}, \citenamefont
  {Wenner}, \citenamefont {White}, \citenamefont {Korotkov}, \citenamefont
  {Cleland},\ and\ \citenamefont {Martinis}}]{Omalley:2015}%
  \BibitemOpen
  \bibfield  {author} {\bibinfo {author} {\bibfnamefont {P.~J.~J.}\
  \bibnamefont {O'Malley}}, \bibinfo {author} {\bibfnamefont {J.}~\bibnamefont
  {Kelly}}, \bibinfo {author} {\bibfnamefont {R.}~\bibnamefont {Barends}},
  \bibinfo {author} {\bibfnamefont {B.}~\bibnamefont {Campbell}}, \bibinfo
  {author} {\bibfnamefont {Y.}~\bibnamefont {Chen}}, \bibinfo {author}
  {\bibfnamefont {Z.}~\bibnamefont {Chen}}, \bibinfo {author} {\bibfnamefont
  {B.}~\bibnamefont {Chiaro}}, \bibinfo {author} {\bibfnamefont
  {A.}~\bibnamefont {Dunsworth}}, \bibinfo {author} {\bibfnamefont {A.~G.}\
  \bibnamefont {Fowler}}, \bibinfo {author} {\bibfnamefont {I.-C.}\
  \bibnamefont {Hoi}}, \bibinfo {author} {\bibfnamefont {E.}~\bibnamefont
  {Jeffrey}}, \bibinfo {author} {\bibfnamefont {A.}~\bibnamefont {Megrant}},
  \bibinfo {author} {\bibfnamefont {J.}~\bibnamefont {Mutus}}, \bibinfo
  {author} {\bibfnamefont {C.}~\bibnamefont {Neill}}, \bibinfo {author}
  {\bibfnamefont {C.}~\bibnamefont {Quintana}}, \bibinfo {author}
  {\bibfnamefont {P.}~\bibnamefont {Roushan}}, \bibinfo {author} {\bibfnamefont
  {D.}~\bibnamefont {Sank}}, \bibinfo {author} {\bibfnamefont {A.}~\bibnamefont
  {Vainsencher}}, \bibinfo {author} {\bibfnamefont {J.}~\bibnamefont {Wenner}},
  \bibinfo {author} {\bibfnamefont {T.~C.}\ \bibnamefont {White}}, \bibinfo
  {author} {\bibfnamefont {A.~N.}\ \bibnamefont {Korotkov}}, \bibinfo {author}
  {\bibfnamefont {A.~N.}\ \bibnamefont {Cleland}},\ and\ \bibinfo {author}
  {\bibfnamefont {J.~M.}\ \bibnamefont {Martinis}},\ }Qubit metrology of
  ultralow phase noise using randomized benchmarking,\ \href
  {https://doi.org/10.1103/PhysRevApplied.3.044009} {\bibfield  {journal}
  {\bibinfo  {journal} {Phys. Rev. Applied}\ }\textbf {\bibinfo {volume} {3}},\
  \bibinfo {pages} {044009} (\bibinfo {year} {2015})}\BibitemShut {NoStop}%
\bibitem [{\citenamefont {Kumar}\ \emph {et~al.}(2016)\citenamefont {Kumar},
  \citenamefont {Sendelbach}, \citenamefont {Beck}, \citenamefont {Freeland},
  \citenamefont {Wang}, \citenamefont {Wang}, \citenamefont {Yu}, \citenamefont
  {Wu}, \citenamefont {Pappas},\ and\ \citenamefont {McDermott}}]{Kumar:2016}%
  \BibitemOpen
  \bibfield  {author} {\bibinfo {author} {\bibfnamefont {P.}~\bibnamefont
  {Kumar}}, \bibinfo {author} {\bibfnamefont {S.}~\bibnamefont {Sendelbach}},
  \bibinfo {author} {\bibfnamefont {M.~A.}\ \bibnamefont {Beck}}, \bibinfo
  {author} {\bibfnamefont {J.~W.}\ \bibnamefont {Freeland}}, \bibinfo {author}
  {\bibfnamefont {Z.}~\bibnamefont {Wang}}, \bibinfo {author} {\bibfnamefont
  {H.}~\bibnamefont {Wang}}, \bibinfo {author} {\bibfnamefont {C.~C.}\
  \bibnamefont {Yu}}, \bibinfo {author} {\bibfnamefont {R.~Q.}\ \bibnamefont
  {Wu}}, \bibinfo {author} {\bibfnamefont {D.~P.}\ \bibnamefont {Pappas}},\
  and\ \bibinfo {author} {\bibfnamefont {R.}~\bibnamefont {McDermott}},\
  }Origin and reduction of $1/f$ magnetic flux noise in superconducting
  devices,\ \href
  {https://doi.org/https://doi.org/10.1103/PhysRevApplied.6.041001} {\bibfield
  {journal} {\bibinfo  {journal} {Phys. Rev. Applied}\ }\textbf {\bibinfo
  {volume} {6}},\ \bibinfo {pages} {041001} (\bibinfo {year}
  {2016})}\BibitemShut {NoStop}%
\bibitem [{\citenamefont {Yan}\ \emph {et~al.}(2016)\citenamefont {Yan},
  \citenamefont {Gustavsson}, \citenamefont {Kamal}, \citenamefont {Birenbaum},
  \citenamefont {Sears}, \citenamefont {Hover}, \citenamefont {Gudmundsen},
  \citenamefont {Rosenberg}, \citenamefont {Samach}, \citenamefont {Weber},
  \citenamefont {Yoder}, \citenamefont {Orlando}, \citenamefont {Clarke},
  \citenamefont {Kerman},\ and\ \citenamefont {Oliver}}]{Yan:2016}%
  \BibitemOpen
  \bibfield  {author} {\bibinfo {author} {\bibfnamefont {F.}~\bibnamefont
  {Yan}}, \bibinfo {author} {\bibfnamefont {S.}~\bibnamefont {Gustavsson}},
  \bibinfo {author} {\bibfnamefont {A.}~\bibnamefont {Kamal}}, \bibinfo
  {author} {\bibfnamefont {J.}~\bibnamefont {Birenbaum}}, \bibinfo {author}
  {\bibfnamefont {A.~P.}\ \bibnamefont {Sears}}, \bibinfo {author}
  {\bibfnamefont {D.}~\bibnamefont {Hover}}, \bibinfo {author} {\bibfnamefont
  {T.~J.}\ \bibnamefont {Gudmundsen}}, \bibinfo {author} {\bibfnamefont
  {D.}~\bibnamefont {Rosenberg}}, \bibinfo {author} {\bibfnamefont
  {G.}~\bibnamefont {Samach}}, \bibinfo {author} {\bibfnamefont
  {S.}~\bibnamefont {Weber}}, \bibinfo {author} {\bibfnamefont {J.~L.}\
  \bibnamefont {Yoder}}, \bibinfo {author} {\bibfnamefont {T.~P.}\ \bibnamefont
  {Orlando}}, \bibinfo {author} {\bibfnamefont {J.}~\bibnamefont {Clarke}},
  \bibinfo {author} {\bibfnamefont {A.~J.}\ \bibnamefont {Kerman}},\ and\
  \bibinfo {author} {\bibfnamefont {W.~D.}\ \bibnamefont {Oliver}},\ }The flux
  qubit revisited to enhance coherence and reproducibility,\ \href
  {https://doi.org/10.1038/ncomms12964} {\bibfield  {journal} {\bibinfo
  {journal} {Nat. Commun.}\ }\textbf {\bibinfo {volume} {7}},\ \bibinfo {pages}
  {12964} (\bibinfo {year} {2016})}\BibitemShut {NoStop}%
\bibitem [{\citenamefont {Hutchings}\ \emph {et~al.}(2017)\citenamefont
  {Hutchings}, \citenamefont {Hertzberg}, \citenamefont {Liu}, \citenamefont
  {Bronn}, \citenamefont {Keefe}, \citenamefont {Brink}, \citenamefont {Chow},\
  and\ \citenamefont {Plourde}}]{Plourde:2017}%
  \BibitemOpen
  \bibfield  {author} {\bibinfo {author} {\bibfnamefont {M.~D.}\ \bibnamefont
  {Hutchings}}, \bibinfo {author} {\bibfnamefont {J.~B.}\ \bibnamefont
  {Hertzberg}}, \bibinfo {author} {\bibfnamefont {Y.}~\bibnamefont {Liu}},
  \bibinfo {author} {\bibfnamefont {N.~T.}\ \bibnamefont {Bronn}}, \bibinfo
  {author} {\bibfnamefont {G.~A.}\ \bibnamefont {Keefe}}, \bibinfo {author}
  {\bibfnamefont {M.}~\bibnamefont {Brink}}, \bibinfo {author} {\bibfnamefont
  {J.~M.}\ \bibnamefont {Chow}},\ and\ \bibinfo {author} {\bibfnamefont
  {B.~L.~T.}\ \bibnamefont {Plourde}},\ }Tunable superconducting qubits with
  flux-independent coherence,\ \href
  {https://doi.org/10.1103/PhysRevApplied.8.044003} {\bibfield  {journal}
  {\bibinfo  {journal} {Phys. Rev. Applied}\ }\textbf {\bibinfo {volume} {8}},\
  \bibinfo {pages} {044003} (\bibinfo {year} {2017})}\BibitemShut {NoStop}%
\bibitem [{\citenamefont {Kou}\ \emph {et~al.}(2017)\citenamefont {Kou},
  \citenamefont {Smith}, \citenamefont {Vool}, \citenamefont {Brierley},
  \citenamefont {Meier}, \citenamefont {Frunzio}, \citenamefont {Girvin},
  \citenamefont {Glazman},\ and\ \citenamefont {Devoret}}]{Kou:2017}%
  \BibitemOpen
  \bibfield  {author} {\bibinfo {author} {\bibfnamefont {A.}~\bibnamefont
  {Kou}}, \bibinfo {author} {\bibfnamefont {W.~C.}\ \bibnamefont {Smith}},
  \bibinfo {author} {\bibfnamefont {U.}~\bibnamefont {Vool}}, \bibinfo {author}
  {\bibfnamefont {R.~T.}\ \bibnamefont {Brierley}}, \bibinfo {author}
  {\bibfnamefont {H.}~\bibnamefont {Meier}}, \bibinfo {author} {\bibfnamefont
  {L.}~\bibnamefont {Frunzio}}, \bibinfo {author} {\bibfnamefont {S.~M.}\
  \bibnamefont {Girvin}}, \bibinfo {author} {\bibfnamefont {L.~I.}\
  \bibnamefont {Glazman}},\ and\ \bibinfo {author} {\bibfnamefont {M.~H.}\
  \bibnamefont {Devoret}},\ }Fluxonium-based artificial molecule with a tunable
  magnetic moment,\ \href {https://doi.org/10.1103/PhysRevX.7.031037}
  {\bibfield  {journal} {\bibinfo  {journal} {Phys. Rev. X}\ }\textbf {\bibinfo
  {volume} {7}},\ \bibinfo {pages} {031037} (\bibinfo {year}
  {2017})}\BibitemShut {NoStop}%
\bibitem [{\citenamefont {Quintana}\ \emph {et~al.}(2017)\citenamefont
  {Quintana}, \citenamefont {Chen}, \citenamefont {Sank}, \citenamefont
  {Petukhov}, \citenamefont {White}, \citenamefont {Kafri}, \citenamefont
  {Chiaro}, \citenamefont {Megrant}, \citenamefont {Barends}, \citenamefont
  {Campbell}, \citenamefont {Chen}, \citenamefont {Dunsworth}, \citenamefont
  {Fowler}, \citenamefont {Graff}, \citenamefont {Jeffrey}, \citenamefont
  {Kelly}, \citenamefont {Lucero}, \citenamefont {Mutus}, \citenamefont
  {Neeley}, \citenamefont {Neill}, \citenamefont {O'Malley}, \citenamefont
  {Roushan}, \citenamefont {Shabani}, \citenamefont {Smelyanskiy},
  \citenamefont {Vainsencher}, \citenamefont {Wenner}, \citenamefont {Neven},\
  and\ \citenamefont {Martinis}}]{Quintana:2017}%
  \BibitemOpen
  \bibfield  {author} {\bibinfo {author} {\bibfnamefont {C.~M.}\ \bibnamefont
  {Quintana}}, \bibinfo {author} {\bibfnamefont {Y.}~\bibnamefont {Chen}},
  \bibinfo {author} {\bibfnamefont {D.}~\bibnamefont {Sank}}, \bibinfo {author}
  {\bibfnamefont {A.~G.}\ \bibnamefont {Petukhov}}, \bibinfo {author}
  {\bibfnamefont {T.~C.}\ \bibnamefont {White}}, \bibinfo {author}
  {\bibfnamefont {D.}~\bibnamefont {Kafri}}, \bibinfo {author} {\bibfnamefont
  {B.}~\bibnamefont {Chiaro}}, \bibinfo {author} {\bibfnamefont
  {A.}~\bibnamefont {Megrant}}, \bibinfo {author} {\bibfnamefont
  {R.}~\bibnamefont {Barends}}, \bibinfo {author} {\bibfnamefont
  {B.}~\bibnamefont {Campbell}}, \bibinfo {author} {\bibfnamefont
  {Z.}~\bibnamefont {Chen}}, \bibinfo {author} {\bibfnamefont {A.}~\bibnamefont
  {Dunsworth}}, \bibinfo {author} {\bibfnamefont {A.~G.}\ \bibnamefont
  {Fowler}}, \bibinfo {author} {\bibfnamefont {R.}~\bibnamefont {Graff}},
  \bibinfo {author} {\bibfnamefont {E.}~\bibnamefont {Jeffrey}}, \bibinfo
  {author} {\bibfnamefont {J.}~\bibnamefont {Kelly}}, \bibinfo {author}
  {\bibfnamefont {E.}~\bibnamefont {Lucero}}, \bibinfo {author} {\bibfnamefont
  {J.~Y.}\ \bibnamefont {Mutus}}, \bibinfo {author} {\bibfnamefont
  {M.}~\bibnamefont {Neeley}}, \bibinfo {author} {\bibfnamefont
  {C.}~\bibnamefont {Neill}}, \bibinfo {author} {\bibfnamefont {P.~J.~J.}\
  \bibnamefont {O'Malley}}, \bibinfo {author} {\bibfnamefont {P.}~\bibnamefont
  {Roushan}}, \bibinfo {author} {\bibfnamefont {A.}~\bibnamefont {Shabani}},
  \bibinfo {author} {\bibfnamefont {V.~N.}\ \bibnamefont {Smelyanskiy}},
  \bibinfo {author} {\bibfnamefont {A.}~\bibnamefont {Vainsencher}}, \bibinfo
  {author} {\bibfnamefont {J.}~\bibnamefont {Wenner}}, \bibinfo {author}
  {\bibfnamefont {H.}~\bibnamefont {Neven}},\ and\ \bibinfo {author}
  {\bibfnamefont {J.~M.}\ \bibnamefont {Martinis}},\ }Observation of
  classical-quantum crossover of $1/f$ flux noise and its paramagnetic
  temperature dependence,\ \href
  {https://doi.org/10.1103/PhysRevLett.118.057702} {\bibfield  {journal}
  {\bibinfo  {journal} {Phys. Rev. Lett.}\ }\textbf {\bibinfo {volume} {118}},\
  \bibinfo {pages} {057702} (\bibinfo {year} {2017})}\BibitemShut {NoStop}%
\bibitem [{\citenamefont {You}\ \emph {et~al.}(2019)\citenamefont {You},
  \citenamefont {Sauls},\ and\ \citenamefont {Koch}}]{You:2019}%
  \BibitemOpen
  \bibfield  {author} {\bibinfo {author} {\bibfnamefont {X.}~\bibnamefont
  {You}}, \bibinfo {author} {\bibfnamefont {J.~A.}\ \bibnamefont {Sauls}},\
  and\ \bibinfo {author} {\bibfnamefont {J.}~\bibnamefont {Koch}},\ }Circuit
  quantization in the presence of time-dependent external flux,\ \href
  {https://doi.org/10.1103/PhysRevB.99.174512} {\bibfield  {journal} {\bibinfo
  {journal} {Phys. Rev. B}\ }\textbf {\bibinfo {volume} {99}},\ \bibinfo
  {pages} {174512} (\bibinfo {year} {2019})}\BibitemShut {NoStop}%
\bibitem [{\citenamefont {Rol}\ \emph {et~al.}(2019)\citenamefont {Rol},
  \citenamefont {Battistel}, \citenamefont {Malinowski}, \citenamefont
  {Bultink}, \citenamefont {Tarasinski}, \citenamefont {Vollmer}, \citenamefont
  {Haider}, \citenamefont {Muthusubramanian}, \citenamefont {Bruno},
  \citenamefont {Terhal},\ and\ \citenamefont {DiCarlo}}]{Rol:2019}%
  \BibitemOpen
  \bibfield  {author} {\bibinfo {author} {\bibfnamefont {M.~A.}\ \bibnamefont
  {Rol}}, \bibinfo {author} {\bibfnamefont {F.}~\bibnamefont {Battistel}},
  \bibinfo {author} {\bibfnamefont {F.~K.}\ \bibnamefont {Malinowski}},
  \bibinfo {author} {\bibfnamefont {C.~C.}\ \bibnamefont {Bultink}}, \bibinfo
  {author} {\bibfnamefont {B.~M.}\ \bibnamefont {Tarasinski}}, \bibinfo
  {author} {\bibfnamefont {R.}~\bibnamefont {Vollmer}}, \bibinfo {author}
  {\bibfnamefont {N.}~\bibnamefont {Haider}}, \bibinfo {author} {\bibfnamefont
  {N.}~\bibnamefont {Muthusubramanian}}, \bibinfo {author} {\bibfnamefont
  {A.}~\bibnamefont {Bruno}}, \bibinfo {author} {\bibfnamefont {B.~M.}\
  \bibnamefont {Terhal}},\ and\ \bibinfo {author} {\bibfnamefont
  {L.}~\bibnamefont {DiCarlo}},\ }Fast, high-fidelity conditional-phase gate
  exploiting leakage interference in weakly anharmonic superconducting qubits,\
  \href {https://doi.org/10.1103/PhysRevLett.123.120502} {\bibfield  {journal}
  {\bibinfo  {journal} {Phys. Rev. Lett.}\ }\textbf {\bibinfo {volume} {123}},\
  \bibinfo {pages} {120502} (\bibinfo {year} {2019})}\BibitemShut {NoStop}%
\bibitem [{\citenamefont {Huang}\ \emph {et~al.}(2021)\citenamefont {Huang},
  \citenamefont {Mundada}, \citenamefont {Gyenis}, \citenamefont {Schuster},
  \citenamefont {Houck},\ and\ \citenamefont {Koch}}]{Huang:2020}%
  \BibitemOpen
  \bibfield  {author} {\bibinfo {author} {\bibfnamefont {Z.}~\bibnamefont
  {Huang}}, \bibinfo {author} {\bibfnamefont {P.~S.}\ \bibnamefont {Mundada}},
  \bibinfo {author} {\bibfnamefont {A.}~\bibnamefont {Gyenis}}, \bibinfo
  {author} {\bibfnamefont {D.~I.}\ \bibnamefont {Schuster}}, \bibinfo {author}
  {\bibfnamefont {A.~A.}\ \bibnamefont {Houck}},\ and\ \bibinfo {author}
  {\bibfnamefont {J.}~\bibnamefont {Koch}},\ }Engineering dynamical sweet spots
  to protect qubits from $1/f$ noise,\ \href
  {https://doi.org/10.1103/PhysRevApplied.15.034065} {\bibfield  {journal}
  {\bibinfo  {journal} {Phys. Rev. Applied}\ }\textbf {\bibinfo {volume}
  {15}},\ \bibinfo {pages} {034065} (\bibinfo {year} {2021})}\BibitemShut
  {NoStop}%
\bibitem [{\citenamefont {Didier}\ \emph {et~al.}(2019)\citenamefont {Didier},
  \citenamefont {Sete}, \citenamefont {Combes},\ and\ \citenamefont
  {da~Silva}}]{Nico:2019}%
  \BibitemOpen
  \bibfield  {author} {\bibinfo {author} {\bibfnamefont {N.}~\bibnamefont
  {Didier}}, \bibinfo {author} {\bibfnamefont {E.~A.}\ \bibnamefont {Sete}},
  \bibinfo {author} {\bibfnamefont {J.}~\bibnamefont {Combes}},\ and\ \bibinfo
  {author} {\bibfnamefont {M.~P.}\ \bibnamefont {da~Silva}},\ }{AC} flux sweet
  spots in parametrically modulated superconducting qubits,\ \href
  {https://doi.org/10.1103/PhysRevApplied.12.054015} {\bibfield  {journal}
  {\bibinfo  {journal} {Phys. Rev. Applied}\ }\textbf {\bibinfo {volume}
  {12}},\ \bibinfo {pages} {054015} (\bibinfo {year} {2019})}\BibitemShut
  {NoStop}%
\bibitem [{\citenamefont {Didier}(2019)}]{Nico:bichro}%
  \BibitemOpen
  \bibfield  {author} {\bibinfo {author} {\bibfnamefont {N.}~\bibnamefont
  {Didier}},\ }Flux control of superconducting qubits at dynamical sweet
  spots,\ \Eprint {https://arxiv.org/abs/1912.09416} {arXiv:1912.09416
  [quant-ph]}  (\bibinfo {year} {2019})\BibitemShut {NoStop}%
\bibitem [{\citenamefont {Bertet}\ \emph {et~al.}(2006)\citenamefont {Bertet},
  \citenamefont {Harmans},\ and\ \citenamefont {Mooij}}]{Bertet:2006}%
  \BibitemOpen
  \bibfield  {author} {\bibinfo {author} {\bibfnamefont {P.}~\bibnamefont
  {Bertet}}, \bibinfo {author} {\bibfnamefont {C.~J. P.~M.}\ \bibnamefont
  {Harmans}},\ and\ \bibinfo {author} {\bibfnamefont {J.~E.}\ \bibnamefont
  {Mooij}},\ }Parametric coupling for superconducting qubits,\ \href
  {https://doi.org/10.1103/PhysRevB.73.064512} {\bibfield  {journal} {\bibinfo
  {journal} {Phys. Rev. B}\ }\textbf {\bibinfo {volume} {73}},\ \bibinfo
  {pages} {064512} (\bibinfo {year} {2006})}\BibitemShut {NoStop}%
\bibitem [{\citenamefont {Niskanen}\ \emph {et~al.}(2007)\citenamefont
  {Niskanen}, \citenamefont {Harrabi}, \citenamefont {Yoshihara}, \citenamefont
  {Nakamura}, \citenamefont {Lloyd},\ and\ \citenamefont
  {Tsai}}]{Niskanen:2007}%
  \BibitemOpen
  \bibfield  {author} {\bibinfo {author} {\bibfnamefont {A.~O.}\ \bibnamefont
  {Niskanen}}, \bibinfo {author} {\bibfnamefont {K.}~\bibnamefont {Harrabi}},
  \bibinfo {author} {\bibfnamefont {F.}~\bibnamefont {Yoshihara}}, \bibinfo
  {author} {\bibfnamefont {Y.}~\bibnamefont {Nakamura}}, \bibinfo {author}
  {\bibfnamefont {S.}~\bibnamefont {Lloyd}},\ and\ \bibinfo {author}
  {\bibfnamefont {J.~S.}\ \bibnamefont {Tsai}},\ }Quantum coherent tunable
  coupling of superconducting qubits,\ \href
  {https://doi.org/10.1126/science.1141324} {\bibfield  {journal} {\bibinfo
  {journal} {Science}\ }\textbf {\bibinfo {volume} {316}},\ \bibinfo {pages}
  {723} (\bibinfo {year} {2007})}\BibitemShut {NoStop}%
\bibitem [{\citenamefont {{Rigetti}}(2009)}]{Rigetti:2009}%
  \BibitemOpen
  \bibfield  {author} {\bibinfo {author} {\bibfnamefont {C.~T.}\ \bibnamefont
  {{Rigetti}}},\ }Quantum gates for superconducting qubits,\ \href
  {https://qulab.eng.yale.edu/documents/theses/Rigetti-PhDthesis-QuantumGatesForSuperconductingQubits-Yale2009.pdf}
  {Ph.D. thesis},\ \bibinfo  {school} {Yale University} (\bibinfo {year}
  {2009})\BibitemShut {NoStop}%
\bibitem [{\citenamefont {Beaudoin}\ \emph {et~al.}(2012)\citenamefont
  {Beaudoin}, \citenamefont {da~Silva}, \citenamefont {Dutton},\ and\
  \citenamefont {Blais}}]{Beaudoin:2012}%
  \BibitemOpen
  \bibfield  {author} {\bibinfo {author} {\bibfnamefont {F.}~\bibnamefont
  {Beaudoin}}, \bibinfo {author} {\bibfnamefont {M.~P.}\ \bibnamefont
  {da~Silva}}, \bibinfo {author} {\bibfnamefont {Z.}~\bibnamefont {Dutton}},\
  and\ \bibinfo {author} {\bibfnamefont {A.}~\bibnamefont {Blais}},\
  }First-order sidebands in circuit {QED} using qubit frequency modulation,\
  \href {https://doi.org/10.1103/PhysRevA.86.022305} {\bibfield  {journal}
  {\bibinfo  {journal} {Phys. Rev. A}\ }\textbf {\bibinfo {volume} {86}},\
  \bibinfo {pages} {022305} (\bibinfo {year} {2012})}\BibitemShut {NoStop}%
\bibitem [{\citenamefont {Strand}\ \emph {et~al.}(2013)\citenamefont {Strand},
  \citenamefont {Ware}, \citenamefont {Beaudoin}, \citenamefont {Ohki},
  \citenamefont {Johnson}, \citenamefont {Blais},\ and\ \citenamefont
  {Plourde}}]{Strand:2013}%
  \BibitemOpen
  \bibfield  {author} {\bibinfo {author} {\bibfnamefont {J.~D.}\ \bibnamefont
  {Strand}}, \bibinfo {author} {\bibfnamefont {M.}~\bibnamefont {Ware}},
  \bibinfo {author} {\bibfnamefont {F.}~\bibnamefont {Beaudoin}}, \bibinfo
  {author} {\bibfnamefont {T.~A.}\ \bibnamefont {Ohki}}, \bibinfo {author}
  {\bibfnamefont {B.~R.}\ \bibnamefont {Johnson}}, \bibinfo {author}
  {\bibfnamefont {A.}~\bibnamefont {Blais}},\ and\ \bibinfo {author}
  {\bibfnamefont {B.~L.~T.}\ \bibnamefont {Plourde}},\ }First-order sideband
  transitions with flux-driven asymmetric transmon qubits,\ \href
  {https://doi.org/10.1103/PhysRevB.87.220505} {\bibfield  {journal} {\bibinfo
  {journal} {Phys. Rev. B}\ }\textbf {\bibinfo {volume} {87}},\ \bibinfo
  {pages} {220505} (\bibinfo {year} {2013})}\BibitemShut {NoStop}%
\bibitem [{\citenamefont {McKay}\ \emph {et~al.}(2016)\citenamefont {McKay},
  \citenamefont {Filipp}, \citenamefont {Mezzacapo}, \citenamefont {Magesan},
  \citenamefont {Chow},\ and\ \citenamefont {Gambetta}}]{McKay:2016}%
  \BibitemOpen
  \bibfield  {author} {\bibinfo {author} {\bibfnamefont {D.~C.}\ \bibnamefont
  {McKay}}, \bibinfo {author} {\bibfnamefont {S.}~\bibnamefont {Filipp}},
  \bibinfo {author} {\bibfnamefont {A.}~\bibnamefont {Mezzacapo}}, \bibinfo
  {author} {\bibfnamefont {E.}~\bibnamefont {Magesan}}, \bibinfo {author}
  {\bibfnamefont {J.~M.}\ \bibnamefont {Chow}},\ and\ \bibinfo {author}
  {\bibfnamefont {J.~M.}\ \bibnamefont {Gambetta}},\ }Universal gate for
  fixed-frequency qubits via a tunable bus,\ \href
  {https://doi.org/10.1103/PhysRevApplied.6.064007} {\bibfield  {journal}
  {\bibinfo  {journal} {Phys. Rev. Applied}\ }\textbf {\bibinfo {volume} {6}},\
  \bibinfo {pages} {064007} (\bibinfo {year} {2016})}\BibitemShut {NoStop}%
\bibitem [{\citenamefont {Naik}\ \emph {et~al.}(2017)\citenamefont {Naik},
  \citenamefont {Leung}, \citenamefont {Chakram}, \citenamefont {Groszkowski},
  \citenamefont {Lu}, \citenamefont {Earnest}, \citenamefont {McKay},
  \citenamefont {Koch},\ and\ \citenamefont {Schuster}}]{Naik:2017}%
  \BibitemOpen
  \bibfield  {author} {\bibinfo {author} {\bibfnamefont {R.~K.}\ \bibnamefont
  {Naik}}, \bibinfo {author} {\bibfnamefont {N.}~\bibnamefont {Leung}},
  \bibinfo {author} {\bibfnamefont {S.}~\bibnamefont {Chakram}}, \bibinfo
  {author} {\bibfnamefont {P.}~\bibnamefont {Groszkowski}}, \bibinfo {author}
  {\bibfnamefont {Y.}~\bibnamefont {Lu}}, \bibinfo {author} {\bibfnamefont
  {N.}~\bibnamefont {Earnest}}, \bibinfo {author} {\bibfnamefont {D.~C.}\
  \bibnamefont {McKay}}, \bibinfo {author} {\bibfnamefont {J.}~\bibnamefont
  {Koch}},\ and\ \bibinfo {author} {\bibfnamefont {D.~I.}\ \bibnamefont
  {Schuster}},\ }Random access quantum information processors using multimode
  circuit quantum electrodynamics,\ \href
  {https://doi.org/https://doi.org/10.1038/s41467-017-02046-6} {\bibfield
  {journal} {\bibinfo  {journal} {Nat. Commun.}\ }\textbf {\bibinfo {volume}
  {8}},\ \bibinfo {pages} {1904} (\bibinfo {year} {2017})}\BibitemShut
  {NoStop}%
\bibitem [{\citenamefont {Roth}\ \emph {et~al.}(2017)\citenamefont {Roth},
  \citenamefont {Ganzhorn}, \citenamefont {Moll}, \citenamefont {Filipp},
  \citenamefont {Salis},\ and\ \citenamefont {Schmidt}}]{Roth:2017}%
  \BibitemOpen
  \bibfield  {author} {\bibinfo {author} {\bibfnamefont {M.}~\bibnamefont
  {Roth}}, \bibinfo {author} {\bibfnamefont {M.}~\bibnamefont {Ganzhorn}},
  \bibinfo {author} {\bibfnamefont {N.}~\bibnamefont {Moll}}, \bibinfo {author}
  {\bibfnamefont {S.}~\bibnamefont {Filipp}}, \bibinfo {author} {\bibfnamefont
  {G.}~\bibnamefont {Salis}},\ and\ \bibinfo {author} {\bibfnamefont
  {S.}~\bibnamefont {Schmidt}},\ }Analysis of a parametrically driven
  exchange-type gate and a two-photon excitation gate between superconducting
  qubits,\ \href {https://doi.org/10.1103/PhysRevA.96.062323} {\bibfield
  {journal} {\bibinfo  {journal} {Phys. Rev. A}\ }\textbf {\bibinfo {volume}
  {96}},\ \bibinfo {pages} {062323} (\bibinfo {year} {2017})}\BibitemShut
  {NoStop}%
\bibitem [{\citenamefont {Mundada}\ \emph {et~al.}(2019)\citenamefont
  {Mundada}, \citenamefont {Zhang}, \citenamefont {Hazard},\ and\ \citenamefont
  {Houck}}]{Mundada:2019}%
  \BibitemOpen
  \bibfield  {author} {\bibinfo {author} {\bibfnamefont {P.}~\bibnamefont
  {Mundada}}, \bibinfo {author} {\bibfnamefont {G.}~\bibnamefont {Zhang}},
  \bibinfo {author} {\bibfnamefont {T.}~\bibnamefont {Hazard}},\ and\ \bibinfo
  {author} {\bibfnamefont {A.}~\bibnamefont {Houck}},\ }Suppression of qubit
  crosstalk in a tunable coupling superconducting circuit,\ \href
  {https://doi.org/10.1103/PhysRevApplied.12.054023} {\bibfield  {journal}
  {\bibinfo  {journal} {Phys. Rev. Applied}\ }\textbf {\bibinfo {volume}
  {12}},\ \bibinfo {pages} {054023} (\bibinfo {year} {2019})}\BibitemShut
  {NoStop}%
\bibitem [{\citenamefont {Reagor}\ \emph {et~al.}(2018)\citenamefont {Reagor},
  \citenamefont {Osborn}, \citenamefont {Tezak}, \citenamefont {Staley},
  \citenamefont {Prawiroatmodjo}, \citenamefont {Scheer}, \citenamefont
  {Alidoust}, \citenamefont {Sete}, \citenamefont {Didier}, \citenamefont
  {da~Silva}, \citenamefont {Acala}, \citenamefont {Angeles}, \citenamefont
  {Bestwick}, \citenamefont {Block}, \citenamefont {Bloom}, \citenamefont
  {Bradley}, \citenamefont {Bui}, \citenamefont {Caldwell}, \citenamefont
  {Capelluto}, \citenamefont {Chilcott}, \citenamefont {Cordova}, \citenamefont
  {Crossman}, \citenamefont {Curtis}, \citenamefont {Deshpande}, \citenamefont
  {El~Bouayadi}, \citenamefont {Girshovich}, \citenamefont {Hong},
  \citenamefont {Hudson}, \citenamefont {Karalekas}, \citenamefont {Kuang},
  \citenamefont {Lenihan}, \citenamefont {Manenti}, \citenamefont {Manning},
  \citenamefont {Marshall}, \citenamefont {Mohan}, \citenamefont
  {O{\textquoteright}Brien}, \citenamefont {Otterbach}, \citenamefont
  {Papageorge}, \citenamefont {Paquette}, \citenamefont {Pelstring},
  \citenamefont {Polloreno}, \citenamefont {Rawat}, \citenamefont {Ryan},
  \citenamefont {Renzas}, \citenamefont {Rubin}, \citenamefont {Russel},
  \citenamefont {Rust}, \citenamefont {Scarabelli}, \citenamefont
  {Selvanayagam}, \citenamefont {Sinclair}, \citenamefont {Smith},
  \citenamefont {Suska}, \citenamefont {To}, \citenamefont {Vahidpour},
  \citenamefont {Vodrahalli}, \citenamefont {Whyland}, \citenamefont {Yadav},
  \citenamefont {Zeng},\ and\ \citenamefont {Rigetti}}]{Matt:2018}%
  \BibitemOpen
  \bibfield  {author} {\bibinfo {author} {\bibfnamefont {M.}~\bibnamefont
  {Reagor}}, \bibinfo {author} {\bibfnamefont {C.~B.}\ \bibnamefont {Osborn}},
  \bibinfo {author} {\bibfnamefont {N.}~\bibnamefont {Tezak}}, \bibinfo
  {author} {\bibfnamefont {A.}~\bibnamefont {Staley}}, \bibinfo {author}
  {\bibfnamefont {G.}~\bibnamefont {Prawiroatmodjo}}, \bibinfo {author}
  {\bibfnamefont {M.}~\bibnamefont {Scheer}}, \bibinfo {author} {\bibfnamefont
  {N.}~\bibnamefont {Alidoust}}, \bibinfo {author} {\bibfnamefont {E.~A.}\
  \bibnamefont {Sete}}, \bibinfo {author} {\bibfnamefont {N.}~\bibnamefont
  {Didier}}, \bibinfo {author} {\bibfnamefont {M.~P.}\ \bibnamefont
  {da~Silva}}, \bibinfo {author} {\bibfnamefont {E.}~\bibnamefont {Acala}},
  \bibinfo {author} {\bibfnamefont {J.}~\bibnamefont {Angeles}}, \bibinfo
  {author} {\bibfnamefont {A.}~\bibnamefont {Bestwick}}, \bibinfo {author}
  {\bibfnamefont {M.}~\bibnamefont {Block}}, \bibinfo {author} {\bibfnamefont
  {B.}~\bibnamefont {Bloom}}, \bibinfo {author} {\bibfnamefont
  {A.}~\bibnamefont {Bradley}}, \bibinfo {author} {\bibfnamefont
  {C.}~\bibnamefont {Bui}}, \bibinfo {author} {\bibfnamefont {S.}~\bibnamefont
  {Caldwell}}, \bibinfo {author} {\bibfnamefont {L.}~\bibnamefont {Capelluto}},
  \bibinfo {author} {\bibfnamefont {R.}~\bibnamefont {Chilcott}}, \bibinfo
  {author} {\bibfnamefont {J.}~\bibnamefont {Cordova}}, \bibinfo {author}
  {\bibfnamefont {G.}~\bibnamefont {Crossman}}, \bibinfo {author}
  {\bibfnamefont {M.}~\bibnamefont {Curtis}}, \bibinfo {author} {\bibfnamefont
  {S.}~\bibnamefont {Deshpande}}, \bibinfo {author} {\bibfnamefont
  {T.}~\bibnamefont {El~Bouayadi}}, \bibinfo {author} {\bibfnamefont
  {D.}~\bibnamefont {Girshovich}}, \bibinfo {author} {\bibfnamefont
  {S.}~\bibnamefont {Hong}}, \bibinfo {author} {\bibfnamefont {A.}~\bibnamefont
  {Hudson}}, \bibinfo {author} {\bibfnamefont {P.}~\bibnamefont {Karalekas}},
  \bibinfo {author} {\bibfnamefont {K.}~\bibnamefont {Kuang}}, \bibinfo
  {author} {\bibfnamefont {M.}~\bibnamefont {Lenihan}}, \bibinfo {author}
  {\bibfnamefont {R.}~\bibnamefont {Manenti}}, \bibinfo {author} {\bibfnamefont
  {T.}~\bibnamefont {Manning}}, \bibinfo {author} {\bibfnamefont
  {J.}~\bibnamefont {Marshall}}, \bibinfo {author} {\bibfnamefont
  {Y.}~\bibnamefont {Mohan}}, \bibinfo {author} {\bibfnamefont
  {W.}~\bibnamefont {O{\textquoteright}Brien}}, \bibinfo {author}
  {\bibfnamefont {J.}~\bibnamefont {Otterbach}}, \bibinfo {author}
  {\bibfnamefont {A.}~\bibnamefont {Papageorge}}, \bibinfo {author}
  {\bibfnamefont {J.-P.}\ \bibnamefont {Paquette}}, \bibinfo {author}
  {\bibfnamefont {M.}~\bibnamefont {Pelstring}}, \bibinfo {author}
  {\bibfnamefont {A.}~\bibnamefont {Polloreno}}, \bibinfo {author}
  {\bibfnamefont {V.}~\bibnamefont {Rawat}}, \bibinfo {author} {\bibfnamefont
  {C.~A.}\ \bibnamefont {Ryan}}, \bibinfo {author} {\bibfnamefont
  {R.}~\bibnamefont {Renzas}}, \bibinfo {author} {\bibfnamefont
  {N.}~\bibnamefont {Rubin}}, \bibinfo {author} {\bibfnamefont
  {D.}~\bibnamefont {Russel}}, \bibinfo {author} {\bibfnamefont
  {M.}~\bibnamefont {Rust}}, \bibinfo {author} {\bibfnamefont {D.}~\bibnamefont
  {Scarabelli}}, \bibinfo {author} {\bibfnamefont {M.}~\bibnamefont
  {Selvanayagam}}, \bibinfo {author} {\bibfnamefont {R.}~\bibnamefont
  {Sinclair}}, \bibinfo {author} {\bibfnamefont {R.}~\bibnamefont {Smith}},
  \bibinfo {author} {\bibfnamefont {M.}~\bibnamefont {Suska}}, \bibinfo
  {author} {\bibfnamefont {T.-W.}\ \bibnamefont {To}}, \bibinfo {author}
  {\bibfnamefont {M.}~\bibnamefont {Vahidpour}}, \bibinfo {author}
  {\bibfnamefont {N.}~\bibnamefont {Vodrahalli}}, \bibinfo {author}
  {\bibfnamefont {T.}~\bibnamefont {Whyland}}, \bibinfo {author} {\bibfnamefont
  {K.}~\bibnamefont {Yadav}}, \bibinfo {author} {\bibfnamefont
  {W.}~\bibnamefont {Zeng}},\ and\ \bibinfo {author} {\bibfnamefont {C.~T.}\
  \bibnamefont {Rigetti}},\ }Demonstration of universal parametric entangling
  gates on a multi-qubit lattice,\ \href
  {https://advances.sciencemag.org/content/4/2/eaao3603} {\bibfield  {journal}
  {\bibinfo  {journal} {Science Advances}\ }\textbf {\bibinfo {volume} {4}}
  (\bibinfo {year} {2018})}\BibitemShut {NoStop}%
\bibitem [{\citenamefont {Caldwell}\ \emph {et~al.}(2018)\citenamefont
  {Caldwell}, \citenamefont {Didier}, \citenamefont {Ryan}, \citenamefont
  {Sete}, \citenamefont {Hudson}, \citenamefont {Karalekas}, \citenamefont
  {Manenti}, \citenamefont {da~Silva}, \citenamefont {Sinclair}, \citenamefont
  {Acala}, \citenamefont {Alidoust}, \citenamefont {Angeles}, \citenamefont
  {Bestwick}, \citenamefont {Block}, \citenamefont {Bloom}, \citenamefont
  {Bradley}, \citenamefont {Bui}, \citenamefont {Capelluto}, \citenamefont
  {Chilcott}, \citenamefont {Cordova}, \citenamefont {Crossman}, \citenamefont
  {Curtis}, \citenamefont {Deshpande}, \citenamefont {Bouayadi}, \citenamefont
  {Girshovich}, \citenamefont {Hong}, \citenamefont {Kuang}, \citenamefont
  {Lenihan}, \citenamefont {Manning}, \citenamefont {Marchenkov}, \citenamefont
  {Marshall}, \citenamefont {Maydra}, \citenamefont {Mohan}, \citenamefont
  {O'Brien}, \citenamefont {Osborn}, \citenamefont {Otterbach}, \citenamefont
  {Papageorge}, \citenamefont {Paquette}, \citenamefont {Pelstring},
  \citenamefont {Polloreno}, \citenamefont {Prawiroatmodjo}, \citenamefont
  {Rawat}, \citenamefont {Reagor}, \citenamefont {Renzas}, \citenamefont
  {Rubin}, \citenamefont {Russell}, \citenamefont {Rust}, \citenamefont
  {Scarabelli}, \citenamefont {Scheer}, \citenamefont {Selvanayagam},
  \citenamefont {Smith}, \citenamefont {Staley}, \citenamefont {Suska},
  \citenamefont {Tezak}, \citenamefont {Thompson}, \citenamefont {To},
  \citenamefont {Vahidpour}, \citenamefont {Vodrahalli}, \citenamefont
  {Whyland}, \citenamefont {Yadav}, \citenamefont {Zeng},\ and\ \citenamefont
  {Rigetti}}]{Shane:2018}%
  \BibitemOpen
  \bibfield  {author} {\bibinfo {author} {\bibfnamefont {S.~A.}\ \bibnamefont
  {Caldwell}}, \bibinfo {author} {\bibfnamefont {N.}~\bibnamefont {Didier}},
  \bibinfo {author} {\bibfnamefont {C.~A.}\ \bibnamefont {Ryan}}, \bibinfo
  {author} {\bibfnamefont {E.~A.}\ \bibnamefont {Sete}}, \bibinfo {author}
  {\bibfnamefont {A.}~\bibnamefont {Hudson}}, \bibinfo {author} {\bibfnamefont
  {P.}~\bibnamefont {Karalekas}}, \bibinfo {author} {\bibfnamefont
  {R.}~\bibnamefont {Manenti}}, \bibinfo {author} {\bibfnamefont {M.~P.}\
  \bibnamefont {da~Silva}}, \bibinfo {author} {\bibfnamefont {R.}~\bibnamefont
  {Sinclair}}, \bibinfo {author} {\bibfnamefont {E.}~\bibnamefont {Acala}},
  \bibinfo {author} {\bibfnamefont {N.}~\bibnamefont {Alidoust}}, \bibinfo
  {author} {\bibfnamefont {J.}~\bibnamefont {Angeles}}, \bibinfo {author}
  {\bibfnamefont {A.}~\bibnamefont {Bestwick}}, \bibinfo {author}
  {\bibfnamefont {M.}~\bibnamefont {Block}}, \bibinfo {author} {\bibfnamefont
  {B.}~\bibnamefont {Bloom}}, \bibinfo {author} {\bibfnamefont
  {A.}~\bibnamefont {Bradley}}, \bibinfo {author} {\bibfnamefont
  {C.}~\bibnamefont {Bui}}, \bibinfo {author} {\bibfnamefont {L.}~\bibnamefont
  {Capelluto}}, \bibinfo {author} {\bibfnamefont {R.}~\bibnamefont {Chilcott}},
  \bibinfo {author} {\bibfnamefont {J.}~\bibnamefont {Cordova}}, \bibinfo
  {author} {\bibfnamefont {G.}~\bibnamefont {Crossman}}, \bibinfo {author}
  {\bibfnamefont {M.}~\bibnamefont {Curtis}}, \bibinfo {author} {\bibfnamefont
  {S.}~\bibnamefont {Deshpande}}, \bibinfo {author} {\bibfnamefont {T.~E.}\
  \bibnamefont {Bouayadi}}, \bibinfo {author} {\bibfnamefont {D.}~\bibnamefont
  {Girshovich}}, \bibinfo {author} {\bibfnamefont {S.}~\bibnamefont {Hong}},
  \bibinfo {author} {\bibfnamefont {K.}~\bibnamefont {Kuang}}, \bibinfo
  {author} {\bibfnamefont {M.}~\bibnamefont {Lenihan}}, \bibinfo {author}
  {\bibfnamefont {T.}~\bibnamefont {Manning}}, \bibinfo {author} {\bibfnamefont
  {A.}~\bibnamefont {Marchenkov}}, \bibinfo {author} {\bibfnamefont
  {J.}~\bibnamefont {Marshall}}, \bibinfo {author} {\bibfnamefont
  {R.}~\bibnamefont {Maydra}}, \bibinfo {author} {\bibfnamefont
  {Y.}~\bibnamefont {Mohan}}, \bibinfo {author} {\bibfnamefont
  {W.}~\bibnamefont {O'Brien}}, \bibinfo {author} {\bibfnamefont
  {C.}~\bibnamefont {Osborn}}, \bibinfo {author} {\bibfnamefont
  {J.}~\bibnamefont {Otterbach}}, \bibinfo {author} {\bibfnamefont
  {A.}~\bibnamefont {Papageorge}}, \bibinfo {author} {\bibfnamefont {J.-P.}\
  \bibnamefont {Paquette}}, \bibinfo {author} {\bibfnamefont {M.}~\bibnamefont
  {Pelstring}}, \bibinfo {author} {\bibfnamefont {A.}~\bibnamefont
  {Polloreno}}, \bibinfo {author} {\bibfnamefont {G.}~\bibnamefont
  {Prawiroatmodjo}}, \bibinfo {author} {\bibfnamefont {V.}~\bibnamefont
  {Rawat}}, \bibinfo {author} {\bibfnamefont {M.}~\bibnamefont {Reagor}},
  \bibinfo {author} {\bibfnamefont {R.}~\bibnamefont {Renzas}}, \bibinfo
  {author} {\bibfnamefont {N.}~\bibnamefont {Rubin}}, \bibinfo {author}
  {\bibfnamefont {D.}~\bibnamefont {Russell}}, \bibinfo {author} {\bibfnamefont
  {M.}~\bibnamefont {Rust}}, \bibinfo {author} {\bibfnamefont {D.}~\bibnamefont
  {Scarabelli}}, \bibinfo {author} {\bibfnamefont {M.}~\bibnamefont {Scheer}},
  \bibinfo {author} {\bibfnamefont {M.}~\bibnamefont {Selvanayagam}}, \bibinfo
  {author} {\bibfnamefont {R.}~\bibnamefont {Smith}}, \bibinfo {author}
  {\bibfnamefont {A.}~\bibnamefont {Staley}}, \bibinfo {author} {\bibfnamefont
  {M.}~\bibnamefont {Suska}}, \bibinfo {author} {\bibfnamefont
  {N.}~\bibnamefont {Tezak}}, \bibinfo {author} {\bibfnamefont {D.~C.}\
  \bibnamefont {Thompson}}, \bibinfo {author} {\bibfnamefont {T.-W.}\
  \bibnamefont {To}}, \bibinfo {author} {\bibfnamefont {M.}~\bibnamefont
  {Vahidpour}}, \bibinfo {author} {\bibfnamefont {N.}~\bibnamefont
  {Vodrahalli}}, \bibinfo {author} {\bibfnamefont {T.}~\bibnamefont {Whyland}},
  \bibinfo {author} {\bibfnamefont {K.}~\bibnamefont {Yadav}}, \bibinfo
  {author} {\bibfnamefont {W.}~\bibnamefont {Zeng}},\ and\ \bibinfo {author}
  {\bibfnamefont {C.}~\bibnamefont {Rigetti}},\ }Parametrically activated
  entangling gates using transmon qubits,\ \href
  {https://doi.org/10.1103/PhysRevApplied.10.034050} {\bibfield  {journal}
  {\bibinfo  {journal} {Phys. Rev. Applied}\ }\textbf {\bibinfo {volume}
  {10}},\ \bibinfo {pages} {034050} (\bibinfo {year} {2018})}\BibitemShut
  {NoStop}%
\bibitem [{\citenamefont {Hong}\ \emph {et~al.}(2020)\citenamefont {Hong},
  \citenamefont {Papageorge}, \citenamefont {Sivarajah}, \citenamefont
  {Crossman}, \citenamefont {Didier}, \citenamefont {Polloreno}, \citenamefont
  {Sete}, \citenamefont {Turkowski}, \citenamefont {da~Silva},\ and\
  \citenamefont {Johnson}}]{Sab:2019}%
  \BibitemOpen
  \bibfield  {author} {\bibinfo {author} {\bibfnamefont {S.~S.}\ \bibnamefont
  {Hong}}, \bibinfo {author} {\bibfnamefont {A.~T.}\ \bibnamefont
  {Papageorge}}, \bibinfo {author} {\bibfnamefont {P.}~\bibnamefont
  {Sivarajah}}, \bibinfo {author} {\bibfnamefont {G.}~\bibnamefont {Crossman}},
  \bibinfo {author} {\bibfnamefont {N.}~\bibnamefont {Didier}}, \bibinfo
  {author} {\bibfnamefont {A.~M.}\ \bibnamefont {Polloreno}}, \bibinfo {author}
  {\bibfnamefont {E.~A.}\ \bibnamefont {Sete}}, \bibinfo {author}
  {\bibfnamefont {S.~W.}\ \bibnamefont {Turkowski}}, \bibinfo {author}
  {\bibfnamefont {M.~P.}\ \bibnamefont {da~Silva}},\ and\ \bibinfo {author}
  {\bibfnamefont {B.~R.}\ \bibnamefont {Johnson}},\ }Demonstration of a
  parametrically activated entangling gate protected from flux noise,\ \href
  {https://doi.org/10.1103/PhysRevA.101.012302} {\bibfield  {journal} {\bibinfo
   {journal} {Phys. Rev. A}\ }\textbf {\bibinfo {volume} {101}},\ \bibinfo
  {pages} {012302} (\bibinfo {year} {2020})}\BibitemShut {NoStop}%
\bibitem [{\citenamefont {Abrams}\ \emph {et~al.}(2020)\citenamefont {Abrams},
  \citenamefont {Didier}, \citenamefont {Johnson}, \citenamefont {Silva},\ and\
  \citenamefont {Ryan}}]{Deanna_2020}%
  \BibitemOpen
  \bibfield  {author} {\bibinfo {author} {\bibfnamefont {D.~M.}\ \bibnamefont
  {Abrams}}, \bibinfo {author} {\bibfnamefont {N.}~\bibnamefont {Didier}},
  \bibinfo {author} {\bibfnamefont {B.~R.}\ \bibnamefont {Johnson}}, \bibinfo
  {author} {\bibfnamefont {M.~P.~d.}\ \bibnamefont {Silva}},\ and\ \bibinfo
  {author} {\bibfnamefont {C.~A.}\ \bibnamefont {Ryan}},\ }Implementation of
  $\rm{XY}$ entangling gates with a single calibrated pulse,\ \href
  {https://doi.org/10.1038/s41928-020-00498-1} {\bibfield  {journal} {\bibinfo
  {journal} {Nature Electronics}\ }\textbf {\bibinfo {volume} {3}},\ \bibinfo
  {pages} {744–750} (\bibinfo {year} {2020})}\BibitemShut {NoStop}%
\bibitem [{\citenamefont {Fried}\ \emph {et~al.}(2019)\citenamefont {Fried},
  \citenamefont {Sivarajah}, \citenamefont {Didier}, \citenamefont {Sete},
  \citenamefont {P.~da Silva}, \citenamefont {Johnson},\ and\ \citenamefont
  {Ryan}}]{Schuyler:2019}%
  \BibitemOpen
  \bibfield  {author} {\bibinfo {author} {\bibfnamefont {E.~S.}\ \bibnamefont
  {Fried}}, \bibinfo {author} {\bibfnamefont {P.}~\bibnamefont {Sivarajah}},
  \bibinfo {author} {\bibfnamefont {N.}~\bibnamefont {Didier}}, \bibinfo
  {author} {\bibfnamefont {E.~A.}\ \bibnamefont {Sete}}, \bibinfo {author}
  {\bibfnamefont {M.}~\bibnamefont {P.~da Silva}}, \bibinfo {author}
  {\bibfnamefont {B.~R.}\ \bibnamefont {Johnson}},\ and\ \bibinfo {author}
  {\bibfnamefont {C.~A.}\ \bibnamefont {Ryan}},\ }Assessing the influence of
  broadband instrumentation noise on parametrically modulated superconducting
  qubits,\ \Eprint {https://arxiv.org/abs/1908.11370} {arXiv:1908.11370}
  (\bibinfo {year} {2019})\BibitemShut {NoStop}%
\bibitem [{\citenamefont {Didier}\ \emph {et~al.}(2018)\citenamefont {Didier},
  \citenamefont {Sete}, \citenamefont {da~Silva},\ and\ \citenamefont
  {Rigetti}}]{Nico:2018}%
  \BibitemOpen
  \bibfield  {author} {\bibinfo {author} {\bibfnamefont {N.}~\bibnamefont
  {Didier}}, \bibinfo {author} {\bibfnamefont {E.~A.}\ \bibnamefont {Sete}},
  \bibinfo {author} {\bibfnamefont {M.~P.}\ \bibnamefont {da~Silva}},\ and\
  \bibinfo {author} {\bibfnamefont {C.}~\bibnamefont {Rigetti}},\ }Analytical
  modeling of parametrically modulated transmon qubits,\ \href
  {https://doi.org/10.1103/PhysRevA.97.022330} {\bibfield  {journal} {\bibinfo
  {journal} {Phys. Rev. A}\ }\textbf {\bibinfo {volume} {97}},\ \bibinfo
  {pages} {022330} (\bibinfo {year} {2018})}\BibitemShut {NoStop}%
\bibitem [{\citenamefont {Koch}\ \emph
  {et~al.}(2007{\natexlab{b}})\citenamefont {Koch}, \citenamefont {Yu},
  \citenamefont {Gambetta}, \citenamefont {Houck}, \citenamefont {Schuster},
  \citenamefont {Majer}, \citenamefont {Blais}, \citenamefont {Devoret},
  \citenamefont {Girvin},\ and\ \citenamefont {Schoelkopf}}]{Koch:2007}%
  \BibitemOpen
  \bibfield  {author} {\bibinfo {author} {\bibfnamefont {J.}~\bibnamefont
  {Koch}}, \bibinfo {author} {\bibfnamefont {T.~M.}\ \bibnamefont {Yu}},
  \bibinfo {author} {\bibfnamefont {J.}~\bibnamefont {Gambetta}}, \bibinfo
  {author} {\bibfnamefont {A.~A.}\ \bibnamefont {Houck}}, \bibinfo {author}
  {\bibfnamefont {D.~I.}\ \bibnamefont {Schuster}}, \bibinfo {author}
  {\bibfnamefont {J.}~\bibnamefont {Majer}}, \bibinfo {author} {\bibfnamefont
  {A.}~\bibnamefont {Blais}}, \bibinfo {author} {\bibfnamefont {M.~H.}\
  \bibnamefont {Devoret}}, \bibinfo {author} {\bibfnamefont {S.~M.}\
  \bibnamefont {Girvin}},\ and\ \bibinfo {author} {\bibfnamefont {R.~J.}\
  \bibnamefont {Schoelkopf}},\ }Charge-insensitive qubit design derived from
  the cooper pair box,\ \href {https://doi.org/10.1103/PhysRevA.76.042319}
  {\bibfield  {journal} {\bibinfo  {journal} {Phys. Rev. A}\ }\textbf {\bibinfo
  {volume} {76}},\ \bibinfo {pages} {042319} (\bibinfo {year}
  {2007}{\natexlab{b}})}\BibitemShut {NoStop}%
\bibitem [{\citenamefont {Krantz}\ \emph {et~al.}(2019)\citenamefont {Krantz},
  \citenamefont {Kjaergaard}, \citenamefont {Yan}, \citenamefont {Orlando},
  \citenamefont {Gustavsson},\ and\ \citenamefont {Oliver}}]{Krantz_2019}%
  \BibitemOpen
  \bibfield  {author} {\bibinfo {author} {\bibfnamefont {P.}~\bibnamefont
  {Krantz}}, \bibinfo {author} {\bibfnamefont {M.}~\bibnamefont {Kjaergaard}},
  \bibinfo {author} {\bibfnamefont {F.}~\bibnamefont {Yan}}, \bibinfo {author}
  {\bibfnamefont {T.~P.}\ \bibnamefont {Orlando}}, \bibinfo {author}
  {\bibfnamefont {S.}~\bibnamefont {Gustavsson}},\ and\ \bibinfo {author}
  {\bibfnamefont {W.~D.}\ \bibnamefont {Oliver}},\ }A quantum engineer’s
  guide to superconducting qubits,\ \href {https://doi.org/10.1063/1.5089550}
  {\bibfield  {journal} {\bibinfo  {journal} {Applied Physics Reviews}\
  }\textbf {\bibinfo {volume} {6}},\ \bibinfo {pages} {021318} (\bibinfo {year}
  {2019})}\BibitemShut {NoStop}%
\bibitem [{\citenamefont {Chuang}\ and\ \citenamefont
  {Nielsen}(1997)}]{Chuang:2009}%
  \BibitemOpen
  \bibfield  {author} {\bibinfo {author} {\bibfnamefont {I.~L.}\ \bibnamefont
  {Chuang}}\ and\ \bibinfo {author} {\bibfnamefont {M.~A.}\ \bibnamefont
  {Nielsen}},\ }Prescription for experimental determination of the dynamics of
  a quantum black box,\ \href {https://doi.org/10.1080/09500349708231894}
  {\bibfield  {journal} {\bibinfo  {journal} {Journal of Modern Optics}\
  }\textbf {\bibinfo {volume} {44}},\ \bibinfo {pages} {2455} (\bibinfo {year}
  {1997})}\BibitemShut {NoStop}%
\bibitem [{\citenamefont {Nielsen}(2002)}]{Nielsen:2002}%
  \BibitemOpen
  \bibfield  {author} {\bibinfo {author} {\bibfnamefont {M.~A.}\ \bibnamefont
  {Nielsen}},\ }A simple formula for the average gate fidelity of a quantum
  dynamical operation,\ \href
  {https://doi.org/https://doi.org/10.1016/S0375-9601(02)01272-0} {\bibfield
  {journal} {\bibinfo  {journal} {Physics Letters A}\ }\textbf {\bibinfo
  {volume} {303}},\ \bibinfo {pages} {249} (\bibinfo {year}
  {2002})}\BibitemShut {NoStop}%
\bibitem [{\citenamefont {Ryan}\ \emph {et~al.}(2015)\citenamefont {Ryan},
  \citenamefont {Johnson}, \citenamefont {Gambetta}, \citenamefont {Chow},
  \citenamefont {da~Silva}, \citenamefont {Dial},\ and\ \citenamefont
  {Ohki}}]{Ryan_2015}%
  \BibitemOpen
  \bibfield  {author} {\bibinfo {author} {\bibfnamefont {C.~A.}\ \bibnamefont
  {Ryan}}, \bibinfo {author} {\bibfnamefont {B.~R.}\ \bibnamefont {Johnson}},
  \bibinfo {author} {\bibfnamefont {J.~M.}\ \bibnamefont {Gambetta}}, \bibinfo
  {author} {\bibfnamefont {J.~M.}\ \bibnamefont {Chow}}, \bibinfo {author}
  {\bibfnamefont {M.~P.}\ \bibnamefont {da~Silva}}, \bibinfo {author}
  {\bibfnamefont {O.~E.}\ \bibnamefont {Dial}},\ and\ \bibinfo {author}
  {\bibfnamefont {T.~A.}\ \bibnamefont {Ohki}},\ }Tomography via correlation of
  noisy measurement records,\ \href
  {https://doi.org/10.1103/PhysRevA.91.022118} {\bibfield  {journal} {\bibinfo
  {journal} {Phys. Rev. A}\ }\textbf {\bibinfo {volume} {91}},\ \bibinfo
  {pages} {022118} (\bibinfo {year} {2015})}\BibitemShut {NoStop}%
\bibitem [{\citenamefont {Knill}\ \emph {et~al.}(2008)\citenamefont {Knill},
  \citenamefont {Leibfried}, \citenamefont {Reichle}, \citenamefont {Britton},
  \citenamefont {Blakestad}, \citenamefont {Jost}, \citenamefont {Langer},
  \citenamefont {Ozeri}, \citenamefont {Seidelin},\ and\ \citenamefont
  {Wineland}}]{Knill_2008}%
  \BibitemOpen
  \bibfield  {author} {\bibinfo {author} {\bibfnamefont {E.}~\bibnamefont
  {Knill}}, \bibinfo {author} {\bibfnamefont {D.}~\bibnamefont {Leibfried}},
  \bibinfo {author} {\bibfnamefont {R.}~\bibnamefont {Reichle}}, \bibinfo
  {author} {\bibfnamefont {J.}~\bibnamefont {Britton}}, \bibinfo {author}
  {\bibfnamefont {R.~B.}\ \bibnamefont {Blakestad}}, \bibinfo {author}
  {\bibfnamefont {J.~D.}\ \bibnamefont {Jost}}, \bibinfo {author}
  {\bibfnamefont {C.}~\bibnamefont {Langer}}, \bibinfo {author} {\bibfnamefont
  {R.}~\bibnamefont {Ozeri}}, \bibinfo {author} {\bibfnamefont
  {S.}~\bibnamefont {Seidelin}},\ and\ \bibinfo {author} {\bibfnamefont
  {D.~J.}\ \bibnamefont {Wineland}},\ }Randomized benchmarking of quantum
  gates,\ \href {https://doi.org/10.1103/PhysRevA.77.012307} {\bibfield
  {journal} {\bibinfo  {journal} {Phys. Rev. A}\ }\textbf {\bibinfo {volume}
  {77}},\ \bibinfo {pages} {012307} (\bibinfo {year} {2008})}\BibitemShut
  {NoStop}%
\bibitem [{\citenamefont {Magesan}\ \emph {et~al.}(2012)\citenamefont
  {Magesan}, \citenamefont {Gambetta}, \citenamefont {Johnson}, \citenamefont
  {Ryan}, \citenamefont {Chow}, \citenamefont {Merkel}, \citenamefont
  {da~Silva}, \citenamefont {Keefe}, \citenamefont {Rothwell}, \citenamefont
  {Ohki}, \citenamefont {Ketchen},\ and\ \citenamefont
  {Steffen}}]{Magesan_2012}%
  \BibitemOpen
  \bibfield  {author} {\bibinfo {author} {\bibfnamefont {E.}~\bibnamefont
  {Magesan}}, \bibinfo {author} {\bibfnamefont {J.~M.}\ \bibnamefont
  {Gambetta}}, \bibinfo {author} {\bibfnamefont {B.~R.}\ \bibnamefont
  {Johnson}}, \bibinfo {author} {\bibfnamefont {C.~A.}\ \bibnamefont {Ryan}},
  \bibinfo {author} {\bibfnamefont {J.~M.}\ \bibnamefont {Chow}}, \bibinfo
  {author} {\bibfnamefont {S.~T.}\ \bibnamefont {Merkel}}, \bibinfo {author}
  {\bibfnamefont {M.~P.}\ \bibnamefont {da~Silva}}, \bibinfo {author}
  {\bibfnamefont {G.~A.}\ \bibnamefont {Keefe}}, \bibinfo {author}
  {\bibfnamefont {M.~B.}\ \bibnamefont {Rothwell}}, \bibinfo {author}
  {\bibfnamefont {T.~A.}\ \bibnamefont {Ohki}}, \bibinfo {author}
  {\bibfnamefont {M.~B.}\ \bibnamefont {Ketchen}},\ and\ \bibinfo {author}
  {\bibfnamefont {M.}~\bibnamefont {Steffen}},\ }Efficient measurement of
  quantum gate error by interleaved randomized benchmarking,\ \href
  {https://doi.org/10.1103/PhysRevLett.109.080505} {\bibfield  {journal}
  {\bibinfo  {journal} {Phys. Rev. Lett.}\ }\textbf {\bibinfo {volume} {109}},\
  \bibinfo {pages} {080505} (\bibinfo {year} {2012})}\BibitemShut {NoStop}%
\bibitem [{\citenamefont {Sete}\ \emph {et~al.}(2021)\citenamefont {Sete},
  \citenamefont {Didier}, \citenamefont {Chen}, \citenamefont {Kulshreshtha},
  \citenamefont {Manenti},\ and\ \citenamefont {Poletto}}]{Sete:2021}%
  \BibitemOpen
  \bibfield  {author} {\bibinfo {author} {\bibfnamefont {E.~A.}\ \bibnamefont
  {Sete}}, \bibinfo {author} {\bibfnamefont {N.}~\bibnamefont {Didier}},
  \bibinfo {author} {\bibfnamefont {A.~Q.}\ \bibnamefont {Chen}}, \bibinfo
  {author} {\bibfnamefont {S.}~\bibnamefont {Kulshreshtha}}, \bibinfo {author}
  {\bibfnamefont {R.}~\bibnamefont {Manenti}},\ and\ \bibinfo {author}
  {\bibfnamefont {S.}~\bibnamefont {Poletto}},\ }Parametric-resonance
  entangling gates with a tunable coupler,\ \Eprint
  {https://arxiv.org/abs/2104.03511} {arXiv:2104.03511 [quant-ph]}  (\bibinfo
  {year} {2021})\BibitemShut {NoStop}%
\end{thebibliography}
\end{document}